\documentclass[a4paper,11pt]{article}
\usepackage{jheppub}
\usepackage[utf8]{inputenc}
\usepackage{graphicx}
\usepackage{tabularx}
\usepackage[dvipsnames]{xcolor}
\usepackage{subcaption}
\usepackage{caption}
\usepackage{amsmath}
\usepackage{amssymb}
\usepackage{enumitem}
\usepackage{tikz}
\usepackage[toc,page]{appendix}
\usepackage{hyperref}

\usepackage[font=footnotesize,labelfont=small]{caption}
\DeclareMathOperator{\Hom}{Hom}

\title{\boldmath Lattice defect networks in 2d Yang-Mills}

\author[a]{Luca Griguolo,}
\author[b]{Elisa Iris Marieni,}
\author[b]{and Itamar Yaakov}
\affiliation[a]{Dipartimento SMFI, Universit\`a di Parma and INFN Gruppo Collegato di Parma\\Viale G.P.\ Usberti 7/A, 43100 Parma, Italy}
\affiliation[b]{STAG Research Center \& Mathematical Sciences, University of Southampton\\Highfield, Southampton SO17 1BJ, UK}

\emailAdd{luca.griguolo@unipr.it}
\emailAdd{e.i.marieni@soton.ac.uk}
\emailAdd{i.yaakov@soton.ac.uk}

\abstract{We construct defect networks in pure Yang-Mills theory in two dimensions using a refinement of the lattice approach. The refinement preserves the locality properties of individual defects, and is compatible with solvability of the theory via subdivision invariance. We explicitly demonstrate closure of the building blocks under fusion.}

\begin{document}
\maketitle
\flushbottom

\section{Introduction}

Defects are spacetime inhomogeneities which generalize the concept of a local operator. Defects have gained increasing importance recently, due to their role in the generalized symmetry program \cite{Kapustin:2014gua,Gaiotto:2014kfa, Bhardwaj:2023kri}. Familiar non-local defects in gauge theory include Wilson loops and surface operators. Despite their ubiquity, we do not know of an algorithm for constructing all the defects of a generic continuum QFT, similar to the state-operator construction for local operators in a CFT. 

 Topological quantum field theories (TQFTs) are one possible exception to the previous statement. TQFTs can reasonably lay claim to being the simplest type of interacting quantum field theory. An axiomatic definition of TQFTs was given by Atiyah in \cite{Atiyah:1989vu}, based on Segal's axiomatization of conformal field theories in two dimensions \cite{segal1988definition}. However, the class of axiomatically defined TQFTs is smaller than the class of all TQFTs defined non-rigorously using e.g. the path integral. Examples of the former include Chern-Simons theory \cite{Witten:1988hf} and Dijkgraff-Witten theory \cite{Dijkgraaf:1989pz}, and an example of the latter is the zero area limit of 2d Yang-Mills \cite{Witten:1991we}.

Full locality is an idea which first arose in the context of TQFTs \cite{Freed:1994ad,Baez:1995xq}. A fully local $d$-dimensional TQFT is expected to assign data to smooth manifolds of all dimensions less than or equal to $d$, as well as to manifolds with boundaries, corners, etc. Cutting and gluing rules allow one to reconstruct bigger manifolds from simple pieces. For example, manifolds of dimension $d-1$ should carry the data of a Hilbert space of states, while the data assigned to $d$ manifolds are the topological invariants given by the partition function.

An axiomatic definition of fully local TQFTs, using the language of higher categories, was given in \cite{Baez:1995xq}. The authors of \cite{Baez:1995xq} also presented a conjecture, the cobordism hypothesis, which gives a structure theorem classifying such models. A somewhat generalized version of this conjecture was proven in \cite{lurie2008classification}. Some examples of fully local topological theories have been constructed
\cite{Freed:1994ad}\footnote{Additional examples are cited in the highly recommended lectures \cite{freed2019lectures}.}. There have also been recent attempts to extend the idea of full locality to all QFTs \cite{grady2021geometric}. The potential benefits, as noted, extend all the way to a complete classification.

In \cite{Kapustin:2010ta}, Kapustin argues that a fully local TQFT implicitly defines a collection of possibly non-local defects. Kapustin argues that, in a fully local TQFT in $d$-dimensions, defects of dimension $k$ should be regarded as the objects of the $k$-category assigned to the $d-k-1$ sphere. Specifically, the $d-1$-category of boundary conditions coincides with the one assigned to a point. The latter plays a central role in the cobordism hypothesis. The same category is related, via the folding trick, to certain defect networks. For an alternative approach to TQFTs with defects using ordinary categories, see \cite{Carqueville:2023jhb} and references therein.

Part of our aim in this paper is to concretely describe this aspect of a particular fully local \textit{non-topological} QFT by explicitly constructing a subset of the defects of the theory. The theory in question is 2d Yang-Mills (2dYM) which, while not quite topological, is exactly solvable by a variety of different methods \cite{Migdal:1975zg,Rusakov:1990rs,Witten:1991we,Witten:1992xu,Blau:1993hj,Sharpe:2014tca, Blau:1995rs}, and arguably much simpler than some of the examples above. In this paper, we will only make use of the lattice approach to 2dYM. We refer the reader to \cite{Griguolo:2024ecw} for a concise introduction to some of the other approaches, and to the review \cite{Cordes:1994fc} for more details.

In the limit of vanishing gauge coupling, equivalently zero area, 2dYM becomes a TQFT, while for generic values of the coupling the theory is only invariant under area-preserving diffeomorphisms. An extension of the concept of full locality to this class of QFTs, which explicitly covers the version of 2dYM with a single coupling constant, was presented in \cite{Runkel2020AreaDependentQF}. We will not make contact with the categorical language used in this reference, and restrict ourselves to the explicit definition and computation of defects. Another attempt at defining fully local 2dYM, based on the BV formalism, was made in \cite{Iraso:2018mwa}. Our approach will be somewhat different and complementary. We will comment on the relationship to the results of \cite{Iraso:2018mwa} in the discussion.

2dYM is known to admit several types of defects. These include ordinary gauge invariant local operators, as well as Wilson lines and Wilson points \cite{Witten:1992xu,Blau:1993hj,Blau:1991mp}. The theory also admits a variety of deformations of the action which preserve its solvability: theta angles, both ordinary and discrete, and deformations of the action by higher Casimirs \cite{Witten:1992xu,Ganor:1994bq,RUSAKOV1994258,Griguolo:2021rke}. We refer to these as $2$-dimensional defects, and treat them accordingly. Nothing substantial changes in our calculations if we add the higher Casimirs terms, and we will not discuss them in detail. There are also networks of defects of different dimensions which are our main source of interest. Depending on the flavor of the theory that one is considering, this list of defects may include redundancies.

Given all that is known about 2dYM, we should explain our motivation for reconsidering its defects on the lattice. The lattice gauge theory approach is believed to be applicable to the study of continuum gauge theories in any dimension. Such theories do not enjoy the solvability properties of 2dYM, nor are their defects easy to classify. Rather than appealing to the special properties of 2dYM, we will refine its lattice model in such a way that the construction of arbitrary defect networks is straightforward. We will then demonstrate compatibility with the known results for 2dYM, including some version of completeness of our list of defects, as a check of the refinement strategy. We believe this approach, which has its own history,\footnote{See \cite{Chen:2024ddr} for a recent interesting discussion and review in this context.} generalizes both to more intricate 2d models and to higher dimensions and, as indicated, is a necessary step towards full locality.

Networks of defects in two-dimensional Yang-Mills theory are also relevant for the maximally supersymmetric $\mathcal{N}=4$ theory in four dimensions via supersymmetric localization. More precisely, a subsector of 1/8 BPS operators of  $\mathcal{N}=4$ SYM was shown in \cite{Pestun:2009nn} to localize to Yang-Mills theory on $\mathbb{S}^2$. It consists of operators supported on a two-sphere embedded into the $\mathbb{R}^4$ spacetime and includes Wilson loops of generic shapes \cite{Drukker:2007qr}, a family of chiral primary operators \cite{Giombi:2009ds} and  ’t Hooft loops linked with the $\mathbb{S}^2$ \cite{Giombi:2009ek}. Their correlations functions are exactly captured by the vacuum expectation value of local and line operators in specific instanton sectors of the 2d theory \cite{Bassetto:1998sr}. These properties have been recently extended \cite{Wang:2020seq} to some defect versions of $\mathcal{N}=4$ SYM theory, obtained by imposing Gaiotto-Witten boundary conditions \cite{Gaiotto:2008sa} on the half-space.

In Section \ref{section: enriching the lattice theory} we describe our proposed enrichement of the lattice model for 2dYM and carry out some basic checks. Section \ref{section: defects}   includes our realization of defect networks and a brief discussion of the relationship to generalized symmetries, additional checks against known results, and a description of the fusion of point-like defects.  A brief review of some aspects of Lie groups and their representations, as well as discrete theta angles, is provided in the appendices.

\section{Enriching the lattice theory}
\label{section: enriching the lattice theory}

2dYM admits a variety of descriptions in terms of fields. We refer the reader to \cite{Witten:1991we,Witten:1992xu}, or the review \cite{Cordes:1994fc} for details.\footnote{We would again highlight the recent interesting analysis in \cite{Iraso:2018mwa} using the BV formalism.} In these continuum versions of the theory, some defects have multiple equivalent descriptions depending on the specific fields used. For instance, Wilson points \cite{Blau:1993hj} can be described as punctures and their associated holonomy, or as ordinary local operators (cf \cite{Beasley:2009mb} section 5).

2dYM also admits a standard lattice gauge theory description. The degrees of freedom are group valued and associated to edges of the lattice. From our defect-centric point of view, the drawback of this description is that some local defects, e.g. Wilson points, are not described as functions of degrees of freedom associated to vertices as one might expect. Although Wilson points admit other lattice descriptions, these descriptions obscure their locality and, in our view, impede the way to full locality.\footnote{However, see \cite{Runkel2020AreaDependentQF} where the authors use the same lattice model with seemingly no impediment.} As noted, these drawbacks are compounded for defect networks and in higher dimensions.

The continuum descriptions point the way towards a partial solution: enrich the description or field content, in this case on the lattice, to accommodate a simpler realization of defects. A reasonable enrichment must retain solvability, in the form of subdivision invariance, while remaining equivalent to the original description in the absence of defects. This section is devoted to enriching the lattice theory for 2dYM by adding enough degrees of freedom to be able to reproduce a large collection of known defects in the form of functionals of those degrees of freedom. We do not claim that this enrichment is unique, only that it does the job.

In section \ref{section: the proposal} we outline our proposal, show that refinement invariance is not spoiled, and comment on the gluing rules. We also check that previously known results, namely the partition function in the absence of defects, possibly including a discrete theta angle, can be obtained from our approach. 

Consider a compact gauge group $G$. The fundamental ingredient that makes the exact solvability of 2d Yang-Mills theory on a lattice possible is the use of Migdal's factor as the action associated to each lattice plaquette \cite{Migdal:1975zg}

\begin{equation}
    \Gamma\left (\mathcal{U}, \epsilon_{w}\right ) = \sum_{\hat{\rho} \in \widehat{G}} \dim\left (\hat{\rho}\right ) \chi_{\hat{\rho}}\left (\mathcal{U}\right ) e^{-\epsilon_{w} c_{2}\left (\hat{\rho}\right )/2} \, ,
    \label{eq: Migdal factor}
\end{equation}

\noindent where:

\begin{itemize}
    \item $\mathcal{U}$ is the holonomy along the edges of a plaquette $w$;
    \item $\epsilon_{w}=e^{2} A_{w}$ is the product of the coupling constant and the area of the plaquette;
    \item $\hat{\rho}$ is a unitary irrep and $\widehat{G}$ is the set of such irreps of $G$;
    \item $\dim\left (\hat{\rho}\right )$ is the dimension of the representation;
    \item $\chi_{\hat{\rho}}\left (\mathcal{U}\right )$ is the character of $\mathcal{U}$ in the representation $\hat{\rho}$;
    \item $c_{2}\left (\hat{\rho}\right )$ is the quadratic Casimir of $\hat{\rho}$.
\end{itemize}
 
\noindent The hallmark of the lattice model for 2dYM is invariance of the Migdal factor under subdivision of a plaquette \cite{Witten:1991we}. The partition function for a surface of genus $g$ is then given by \cite{Witten:1991we}

\begin{equation}
    Z_{\Sigma_{g}} \left (\epsilon\right ) = \sum_{\hat{\rho} \in \widehat{G}} \frac{e^{-\epsilon c_{2}\left (\hat{\rho}\right )/2}}{\dim\left (\hat{\rho}\right )^{2g - 2}} \, .
    \label{eq: pf witten}
\end{equation}
Due to subdivision invariance, this expression can be derived from a presentation using a single plaquette. It can also be computed by gluing together any number of plaquettes using a standard gluing prescription. 

\subsection{The proposal}
\label{section: the proposal}

\begin{figure}
    \centering
    \includegraphics[scale=0.8]{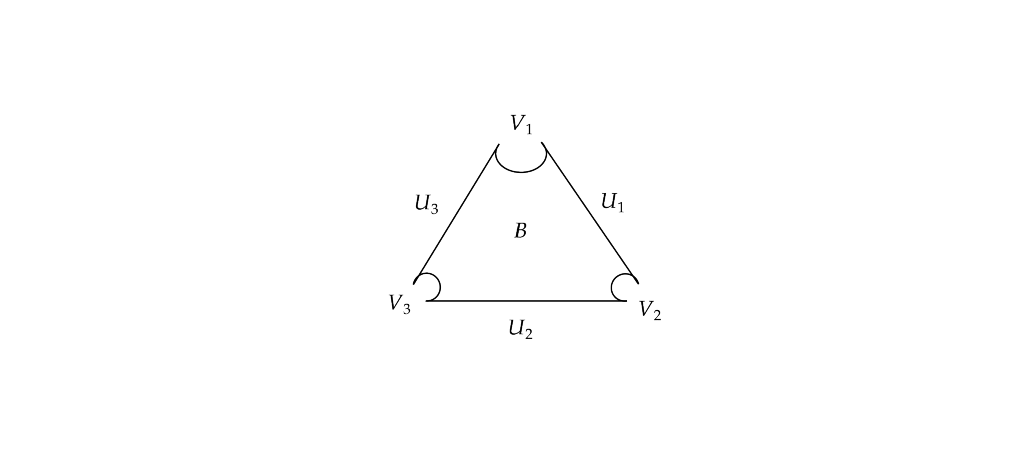}
    \caption{We can recover any orientable surface by gluing together triangular plaquettes like these. We pictorially represent the degrees of freedom assigned to the corners as bites taken out of the plaquette.}
    \label{fig: fundam building block}
\end{figure}

We now reconsider the lattice model for 2dYM, restricting ourselves to a simple, connected and compact gauge group $G = \tilde{G}/H$, with $\tilde{G}$ being the universal cover of $G$, and $H$ a discrete subgroup of the center of $\tilde{G}$. We will also restrict ourselves to orientable surfaces. 
We would like to refine the lattice model to accommodate defects supported on the corners and in the interior of a plaquette. The data assigned to each plaquette will be more complicated than group elements assigned to edges. We take it to consist of:

\begin{enumerate}
    \item corners: we assign group elements $V \in \Tilde{G}$ to corners. Wilson points will appear as functionals of $V$ degrees of freedom;
    \item edges: we assign the standard group elements $U \in \Tilde{G}$ to edges. Wilson lines will appear as functionals of $U$ degrees of freedom;
    \item interior: we assign elements of the center $B \in H$ to plaquettes. Discrete theta angles will appear as functionals of $B$.
\end{enumerate}

The degrees of freedom in points 1 and 3 differ from the standard description of lattice gauge theory, but are not completely novel. The assignment of the corner degrees of freedom is a variation on the central idea in \cite{Iraso:2018mwa}, but now at the group rather than the algebra level. The plaquette degrees of freedom have been considered e.g. in \cite{Santilli:2024dyz}. We will demonstrate that this enriched lattice model is sufficient to reproduce a large class of defect networks as functionals of the lattice degrees of freedom.

We propose the following Migdal's factor for the plaquette enriched with this data (see figure \ref{fig: fundam building block})

\begin{equation}
    \Gamma \left ( \mathcal{U}, B; \epsilon_{w}\right ) = \sum_{\hat{\rho} \in \widehat{\Tilde{G}}} \dim\left (\hat{\rho}\right ) \chi_{\hat{\rho}}\left ( V_{1}U_{1}V_{2}U_{2}V_{3}U_{3}\right ) \chi_{\hat{h}_{\hat{\rho}}}\left( B\right ) e^{-\epsilon_{w} c_{2}\left (\hat{\rho}\right )/2} \, ,
    \label{eq: new Migdal's factor}
\end{equation}

\noindent where,

\begin{itemize}
    \item $U_{1}V_{1}...U_{3}$ is the holonomy along the edges of the plaquette $w$;
    \item $\chi_\cdot$ indicates the character of a group element in a given representation.
    \item $B \in H$ is a discrete degree of freedom associated to the interior of the plaquette;
    \item $\epsilon_{w}=e^{2} A_{w}$ is the product of the coupling constant and the area of the plaquette;
    \item $\hat{\rho}$ is a unitary irrep and $\widehat{\tilde{G}}$ is the set of such irreps;
    \item $\dim\left (\hat{\rho}\right )$ is the dimension of the representation;
    \item $c_{2}\left (\hat{\rho}\right )$ is the quadratic Casimir of $\hat{\rho}$;
    \item $\hat{h}_{\hat{\rho}}$ is the representation of $H$ associated to $\hat{\rho}$ \footnote{We will refer to this representation as the "$N$-ality" of $\hat{\rho}$ even when the group in question is not $SU(N)$.}. We denote by $\hat e$ the trivial representation.
\end{itemize}
Note that we are no longer summing over representations of $G$, but of $\Tilde{G}$. This enables us, for instance, to recover a plaquette with a discrete theta angle, as we will show in the following.\par

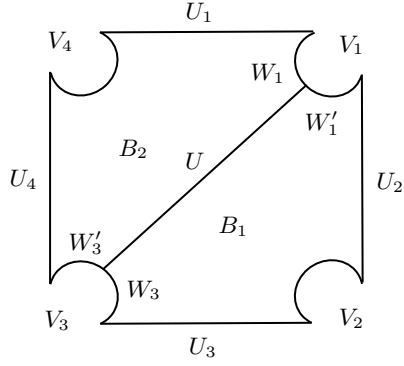
\begin{figure}
    \centering
    \tikzset{every picture/.style={line width=0.75pt}}

\tikzset{every picture/.style={line width=0.75pt}}

\begin{tikzpicture}[x=0.75pt,y=0.75pt,yscale=-1,xscale=1]

\draw    (139.38,60.97) -- (139.38,167.64) ;

\draw    (164.45,41.1) -- (270.76,40.61) ;

\draw    (164.45,187.67) -- (269.94,187.18) ;

\draw    (295.01,62.27) -- (295.39,166.83) ;

\draw  [draw opacity=0] (164.18,41.03) .. controls (166.01,41.83) and (167.69,43) .. (169.11,44.53) .. controls (175.07,50.97) and (174.12,61.46) .. (166.98,67.96) .. controls (159.83,74.45) and (149.21,74.49) .. (143.25,68.05) .. controls (141.49,66.14) and (140.33,63.89) .. (139.76,61.48) -- (156.18,56.29) -- cycle ; \draw   (164.18,41.03) .. controls (166.01,41.83) and (167.69,43) .. (169.11,44.53) .. controls (175.07,50.97) and (174.12,61.46) .. (166.98,67.96) .. controls (159.83,74.45) and (149.21,74.49) .. (143.25,68.05) .. controls (141.49,66.14) and (140.33,63.89) .. (139.76,61.48) ;  

\draw  [draw opacity=0] (294.77,62.93) .. controls (294.19,64.83) and (293.22,66.62) .. (291.85,68.2) .. controls (286.1,74.83) and (275.48,75.12) .. (268.14,68.86) .. controls (260.79,62.59) and (259.5,52.14) .. (265.25,45.51) .. controls (266.95,43.55) and (269.08,42.14) .. (271.42,41.3) -- (278.55,56.86) -- cycle ; \draw   (294.77,62.93) .. controls (294.19,64.83) and (293.22,66.62) .. (291.85,68.2) .. controls (286.1,74.83) and (275.48,75.12) .. (268.14,68.86) .. controls (260.79,62.59) and (259.5,52.14) .. (265.25,45.51) .. controls (266.95,43.55) and (269.08,42.14) .. (271.42,41.3) ;  

\draw  [draw opacity=0] (139.54,167.81) .. controls (139.98,165.87) and (140.82,164.01) .. (142.07,162.34) .. controls (147.31,155.3) and (157.88,154.24) .. (165.67,159.95) .. controls (173.46,165.66) and (175.52,175.99) .. (170.28,183.02) .. controls (168.73,185.1) and (166.71,186.66) .. (164.44,187.68) -- (156.17,172.68) -- cycle ; \draw   (139.54,167.81) .. controls (139.98,165.87) and (140.82,164.01) .. (142.07,162.34) .. controls (147.31,155.3) and (157.88,154.24) .. (165.67,159.95) .. controls (173.46,165.66) and (175.52,175.99) .. (170.28,183.02) .. controls (168.73,185.1) and (166.71,186.66) .. (164.44,187.68) ;  

\draw  [draw opacity=0] (269.94,187.19) .. controls (268.14,186.32) and (266.5,185.1) .. (265.13,183.52) .. controls (259.41,176.87) and (260.74,166.42) .. (268.1,160.18) .. controls (275.47,153.94) and (286.09,154.27) .. (291.82,160.92) .. controls (293.51,162.88) and (294.59,165.18) .. (295.07,167.61) -- (278.48,172.22) -- cycle ; \draw   (269.94,187.19) .. controls (268.14,186.32) and (266.5,185.1) .. (265.13,183.52) .. controls (259.41,176.87) and (260.74,166.42) .. (268.1,160.18) .. controls (275.47,153.94) and (286.09,154.27) .. (291.82,160.92) .. controls (293.51,162.88) and (294.59,165.18) .. (295.07,167.61) ;  

\draw    (166.09,159.99) -- (267.47,67.48) ;

\draw (205.25,99.92) node [anchor=north west][inner sep=0.75pt]  [font=\footnotesize]  {$U$};

\draw (205.54,24.82) node [anchor=north west][inner sep=0.75pt]  [font=\footnotesize]  {$U_{1}$};

\draw (300.79,110.32) node [anchor=north west][inner sep=0.75pt]  [font=\footnotesize]  {$U_{2}$};

\draw (207.18,193.38) node [anchor=north west][inner sep=0.75pt]  [font=\footnotesize]  {$U_{3}$};

\draw (117.68,108.69) node [anchor=north west][inner sep=0.75pt]  [font=\footnotesize]  {$U_{4}$};

\draw (281.99,41.11) node [anchor=north west][inner sep=0.75pt]  [font=\footnotesize]  {$V_{1}$};

\draw (281.99,178.72) node [anchor=north west][inner sep=0.75pt]  [font=\footnotesize]  {$V_{2}$};

\draw (135.01,179.54) node [anchor=north west][inner sep=0.75pt]  [font=\footnotesize]  {$V_{3}$};

\draw (136.65,39.48) node [anchor=north west][inner sep=0.75pt]  [font=\footnotesize]  {$V_{4}$};

\draw (265,79.4) node [anchor=north west][inner sep=0.75pt]  [font=\footnotesize]  {$W'_{1}$};

\draw (238,55.4) node [anchor=north west][inner sep=0.75pt]  [font=\footnotesize]  {$W_{1}$};

\draw (176,164.4) node [anchor=north west][inner sep=0.75pt]  [font=\footnotesize]  {$W_{3}$};

\draw (147,139.4) node [anchor=north west][inner sep=0.75pt]  [font=\footnotesize]  {$W'_{3}$};

\draw (222,132.4) node [anchor=north west][inner sep=0.75pt]  [font=\footnotesize]  {$B_{1}$};

\draw (172,93.4) node [anchor=north west][inner sep=0.75pt]  [font=\footnotesize]  {$B_{2}$};

\end{tikzpicture}

    \caption{Subdivision of a square into two triangles. The holonomy around the square plaquette is $U_{1}V_{1}U_{2}V_{2}U_{3}V_{3}U_{4}V_{4}$. The additional $U$ degree of freedom splits $V_{1},V_{3}$ into $W_{1},W'_{1}$ and $W_{3},W'_{3}$ respectively, so that these are contained in the holonomies around the triangles. $B_{1},B_{2}$ are the discrete degrees of freedom associated to the two triangles, $B$ is the one associated to the square but it does not appear in the picture.}
    \label{fig: refinement invariance}
\end{figure}

When subdividing a plaquette, we would like to retain this data. Such a subdivision is depicted in figure \ref{fig: refinement invariance}. We have indicated, pictorially, the degrees of freedom in the subdivided lattice which we will momentarily integrate out. Subdivision invariance is the statement that, in the absence of any defects associated with the internal degrees of freedom in a plaquette, integrating these out gives back the original plaquette.

Let us call $U$ the additional degree of freedom in the subdivided plaquette. Recall that $U$ is a continuous degree of freedom integrated over the compact group manifold $\tilde{G}$ using the Haar measure, as reviewed in Appendix \ref{appendix: harmonic analysis}. $B$ is a discrete degree of freedom and is summed over its entire range with weight given by  $1/|H|$ to recover the same normalization as the Haar measure. Migdal's factor associated to the square plaquette is

\begin{equation}
    \Gamma_{\text{square}} \left (\mathcal{U}, B; \epsilon\right ) = \sum_{\hat{\rho} \in \widehat{\Tilde{G}}} \dim\left (\hat{\rho}\right ) \chi_{\hat{\rho}}\left (U_{1}V_{1}U_{2}V_{2}U_{3}V_{3}U_{4}V_{4}\right ) \chi_{\hat{h}_{\hat{\rho}}}\left (B\right ) e^{-\epsilon c_{2}\left (\hat{\rho}\right )/2} \, .
\end{equation}
Subdividing this plaquette means adding an internal degree of freedom $U$, the product of the factors associated to the resulting triangles is 

\begin{equation}
\begin{split}
    \Gamma_{1} \left (\mathcal{U}_{1}, B_{1}; \epsilon_{1}\right ) \Gamma_{2} \left (\mathcal{U}_{2}, B_{2}; \epsilon_{2}\right ) =& \sum_{\hat{\rho}, \hat{\sigma} \in \widehat{\Tilde{G}}} \dim\left (\hat{\rho}\right ) \dim\left (\hat{\sigma}\right ) \chi_{\hat{\rho}} \left (W'_{1}U_{2}V_{2}U_{3}W_{3} U\right ) \chi_{\hat{h}_{\hat{\rho}}}\left (B_{1}\right ) \times \\
    &\times \chi_{\hat{\sigma}}\left (U^{-1}W'_{3}U_{4}V_{4}U_{1}W_{1}\right ) \chi_{\hat{h}_{\hat{\sigma}}}\left (B_{2}\right ) e^{-\epsilon_{1}c_{2}\left (\hat{\rho}\right )/2 - \epsilon_{2}c_{2}\left (\hat{\sigma}\right )/2} \, .
\end{split}
\end{equation}
Using (\ref{eq: schur orthogonality}) one can show that $\int dU \Gamma_{1} \Gamma_{2} = \Gamma$. The square plaquette depends on four $U$s, four $V$s, $B$ and $\epsilon$. What we showed proves that when adding an internal degree of freedom the resulting object retains the same amount of data and has the same action. Indeed, after integrating out the internal degree of freedom the resulting plaquette no longer depends on $W_{1}$ and $W'_{1}$ separately, but on the combination $W_{1}W'_{1}$ that we can rename $V_{1}$. The same is true for $W_{3}W_{3}' = V_{3}$, for $\epsilon_{1} + \epsilon_{2} = \epsilon$ and for $B_{1}B_{2} = B$.

Above, we used the standard gluing rules described in e.g.  \cite{Witten:1991we}:. Specifically, we integrated out all the internal degrees of freedom with a weight given by the Haar measure and our choice of normalization.
Changing this weight appropriately, we can recover various defects. For instance, integrating out a loop of  $U$ 'swith a character as weight we recover a Wilson loop. We can also demonstrate that adding a weight when integrating out the $B$'s enables one to recover discrete theta angles  \cite{Santilli:2024dyz}, interpreted here as space-filling defects.
Below, we make some basic checks of our lattic model. In section \ref{section: defects}, we will discuss the more general weights needed to recover defect networks.

\paragraph{Recovering the partition function}
\label{section: recovering PF}

We can recover the partition function (\ref{eq: pf witten}) just by integrating out all degrees of freedom. Let us consider the simple example of a 2-sphere obtained by gluing together two triangular plaquettes

\begin{equation}
    \begin{split}
    Z_{\mathbb{S}^{2}}\left (\epsilon\right ) =& \frac{1}{|H|^{2}} \int dU_{i} dV_{i} \sum_{B_{1},B_{2}  \in H}  \sum_{\hat{\sigma}, \hat{\rho} \in \widehat{\tilde{G}}} \dim \left (\hat{\sigma} \right ) \dim \left (\hat{\rho} \right )\, \chi_{\hat{h}_{\hat{\rho}}}\left (B_{1}\right ) \chi_{\hat{h}_{\hat{\sigma}}}\left (B_{2}\right ) \times \\ 
    &\times \chi_{\hat{\rho}}\left (U_{1}V_{1}U_{2}V_{2}U_{3}V_{3}\right ) \chi_{\hat{\sigma}}\left (U_{1}^{-1}V_{1}^{-1}U_{3}^{-1}V_{3}^{-1}U_{2}^{-1}V_{2}^{-1}\right )
     e^{-\epsilon_{1}c_{2}\left (\hat{\sigma}\right )/2 -\epsilon_{2}c_{2}\left (\hat{\rho}\right )/2}\\
    =& \sum_{\hat{\sigma} \in \widehat{\tilde{G}}} \left (\dim \hat{\sigma}\right )^{2} \, \delta_{\hat e, \hat{h}_{\hat{\sigma}}} \,  e^{-\epsilon c_{2}\left (\hat{\sigma}\right )/2}\\
    =& \sum_{\hat{\sigma} \in \widehat{G}} \left (\dim \hat{\sigma}\right )^{2}  e^{-\epsilon c_{2}\left (\hat{\sigma}\right )/2} \, ,
    \end{split}
\end{equation}
which is indeed the partition function for the $G$ theory.

\paragraph{Recovering the partition function with a discrete theta angle}
\label{section: recovering theta}

One may also consider the partition function of 2dYM with a discrete theta angle for $G$ \cite{Sharpe:2014tca,Santilli:2024dyz} (see Appendix \ref{appendix: theta terms}).
We view such theta angles as space-filling defects with data $\hat{\theta} \in \widehat{H}$. The modified Migdal's factor (\ref{eq: new Migdal's factor}) enables us to reproduce them by integrating out the degree of freedom $B$ using the following prescription \par

\begin{equation}
    \frac{1}{|H|} \sum_{B \in H} \chi_{\hat{\theta}}\left (B^{-1}\right ) \, ,
    \label{eq: integrate out B}
\end{equation}
\noindent to obtain 
\begin{equation}
    \Gamma \left (\mathcal{U}, \epsilon_{w}, \hat{\theta}\right ) = \sum_{\hat{\rho} \in \widehat{\Tilde{G}}} \dim \left (\hat{\rho}\right ) \, \delta_{\hat{h}_{\hat{\rho}}, \hat{\theta}} \, \chi_{\hat{\rho}} \left (\mathcal{U}\right ) e^{-\epsilon_{w} c_{2} \left (\hat{\rho}\right ) / 2} \, .
    \label{eq: Migdal's factor with theta angle}
\end{equation}
 $B \in H$ is a discrete degree of freedom whose interpretation is as the background gauge field associated to the center 1-form symmetry $H^{\left (1\right )} \subseteq Z\left (\tilde{G}\right )^{\left (1\right )}$ of the $\tilde{G}$ theory \footnote{The superscript $\left (1\right )$ indicates the degree of the generalized symmetry.}. It is a discrete analog of a 2-form gauge field. Making it dynamical amounts to gauging the discrete 1-form symmetry and projecting to a $G$ theory (cf eq. 3.18 in \cite{Lin:2022xod}). 
 
 Representations of $\tilde{G}$ split into a number of copies of representations of $G$ labeled by the $N$-ality: the choice of a theta angle amounts to choosing one of these copies.  A discrete theta angle is a sum over irreducible representations of the universal cover $\hat{\rho} \in \Tilde{G}$, with the restriction $\delta_{\hat{h}_{\hat{\rho}}, \hat{\theta}}$. As shown in the previous paragraph, if we choose the theta angle in the trivial representation $\hat{\theta} = \hat e$ we are summing representations $\hat{\rho} \in \widehat{G}$ and recover the original partition function. Instead, the partition function for a surface of genus $g$ with an arbitrary discrete theta angle is given by

\begin{equation}
    Z_{\Sigma_{g}}\left (\epsilon; \hat{\theta}\right ) = \sum_{\hat{\rho} \in \widehat{\Tilde{G}}} \dim \left (\hat{\rho}\right )^{2-2g} \, \delta_{\hat{h}_{\hat{\rho}},\hat{\theta}} \, e^{-\epsilon c_{2}\left (\hat{\rho}\right )/2} \, .
    \label{eq: pf with theta}
\end{equation}

This type of expression appears in \cite{Santilli:2024dyz} as a result of gauging the $SU\left (N\right )$ theory's 1-form symmetry. To prove (\ref{eq: pf with theta}) we follow the same proof given in \cite{Witten:1991we} for (\ref{eq: pf witten}). Suppose we have a genus $g>1$ surface. We can decompose it in $2g-2$ 3-holed spheres. Let us call $U,W,V$ the holonomies around the boundaries and $\hat{\alpha}, \hat{\beta}, \hat{\gamma}$ their representations. Using the new basis, for each 3-holed sphere $\Sigma$ we get

\begin{equation}
\begin{split}
    Z_{\Sigma}\left (\epsilon_{\Sigma}, \hat{\alpha}, \hat{\theta}\right ) =& \sum_{\hat{\sigma}  \in \widehat{\Tilde{G}}}  \int dU dV dW dX dY \, \delta_{\hat{h}_{\hat{\sigma}}, \hat{\theta}} \, \chi_{\hat{\sigma}} \left (X^{-1}UXVY^{-1}WY\right ) \times \\
    & \times \overline{\chi_{\hat{\alpha}}\left (U\right )} \, \overline{\chi_{\hat{\beta}}\left (V\right )} \, \overline{\chi_{\hat{\gamma}}\left (W\right )} \,\dim \left (\hat{\sigma}\right ) e^{-\epsilon_{\Sigma} c_{2}\left (\hat{\sigma}\right )/2} \\
=& \delta_{\hat{h}_{\hat{\alpha}}, \hat{\theta}} \, \delta_{\hat{\alpha}, \hat{\beta}, \hat{\gamma}} \, \frac{e^{-\epsilon_{\Sigma} c_{2}\left (\hat{\alpha}\right )/2}}{\dim\left (\hat{\alpha}\right )} \, .
\end{split}
\end{equation}
Taking the product of the $2g-2$ pieces we recover (\ref{eq: pf with theta}).

\paragraph{Decomposition}
The partition function for the $\tilde{G}$ theory manifestly decomposes into that of different $\tilde{G}/H$ theories, labeled by $\hat{\theta}$

\begin{equation}
    Z_{\Sigma_{g}}\left (\tilde{G}\right ) = \frac{1}{|H|} \sum_{\hat{\theta} \in \widehat{H}} Z_{\Sigma_{g}} \left (G;\hat{\theta}\right ) = \frac{1}{|H|} \sum_{\hat{\theta} \in \widehat{H}} \sum_{\hat{\rho} \in \widehat{\Tilde{G}}} \dim \left (\hat{\rho}\right )^{2-2g} \, \delta_{\hat{h}_{\hat{\rho}},\hat{\theta}} \, e^{-\epsilon c_{2}\left (\hat{\rho}\right )/2} \, .
\end{equation}
In fact, it is possible to have a finer decomposition. One can define a generalized theta term and show that 2dYM decomposes as an infinite sum of theories labeled by representations of the gauge group. This generalized theta term arises from the gauging of the full non-invertible 1-form symmetry. See \cite{Sharpe:2022ene} for a general review, \cite{Nguyen:2021naa} or section 5 in \cite{Santilli:2024dyz} for the specific application to 2dYM.

\section{Defects}
\label{section: defects}

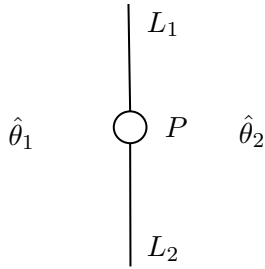
\begin{figure}
    \centering
\tikzset{every picture/.style={line width=0.75pt}}

\begin{tikzpicture}[x=0.75pt,y=0.75pt,yscale=-1,xscale=1]

\draw   (179,119) .. controls (179,114.58) and (182.66,111) .. (187.17,111) .. controls (191.68,111) and (195.33,114.58) .. (195.33,119) .. controls (195.33,123.42) and (191.68,127) .. (187.17,127) .. controls (182.66,127) and (179,123.42) .. (179,119) -- cycle ;

\draw    (187.33,189) -- (187.17,127) ;

\draw    (186.33,57) -- (187.17,111) ;

\draw (125,112.4) node [anchor=north west][inner sep=0.75pt]    {$\hat{\theta }_{1}$};

\draw (240,111.4) node [anchor=north west][inner sep=0.75pt]    {$\hat{\theta }_{2}$};

\draw (194,59.4) node [anchor=north west][inner sep=0.75pt]    {$L_{1}$};

\draw (194,172.4) node [anchor=north west][inner sep=0.75pt]    {$L_{2}$};

\draw (203,112.4) node [anchor=north west][inner sep=0.75pt]    {$P$};

\end{tikzpicture}
 \caption{A defect network incorporating defects of all co-dimensions.}
    \label{fig: network of defects}
\end{figure}

In section \ref{section: the proposal} we enriched the lattice gauge theory version of 2dYM with new data associated to every plaquette and prescribed gluing rules. We showed that this version of the theory could be used to recover the partition function, including possible discrete theta angles. 

We would now like to demonstrate that we can also compute the expectation value of defect networks. Specifically, we would like to incorporate point defects in the interior of a Wilson line, and Wilson lines separating different discrete theta angles, as in figure \ref{fig: network of defects}. We deal with this by prescribing a new systematic way of engineering all defects in terms of building blocks that we will call n-line junctions $\mathbb{W}_{n}$ (left hand side of figure \ref{fig: n-line junction}). \par
In section \ref{section: elementary defects} we define the n-line junction. We explicitly show the 3-line junction that plays the central role in our description and the 4-line junction that we use in some computations. We then outline a few operations that can be performed on the junctions: composition, fusion and degeneration (i.e. erasing lines). In this section, we also state how to build the most general network of defects and comment on the appearance of selection rules. In section \ref{section: degeneration} we show how ($n<3$)-line junctions can be obtained as degenerate cases of 3-line junctions by erasing lines. Here we also show how to recover Wilson lines and Wilson points in our setup. In section \ref{section: fusion} we show that we can recover ($n>3$)-line junctions by fusing 3-line junctions, thus proving that our set of defects is closed under fusion and can be completely reconstructed by the fundamental building block $\mathbb{W}_{3}$. Finally, in section \ref{section: IRF crossing} we show that we can also recover the "IRF crossing": the simplest network of intersecting Wilson lines described in \cite{Witten:1991we}.

\tikzset{every picture/.style={line width=0.75pt}}

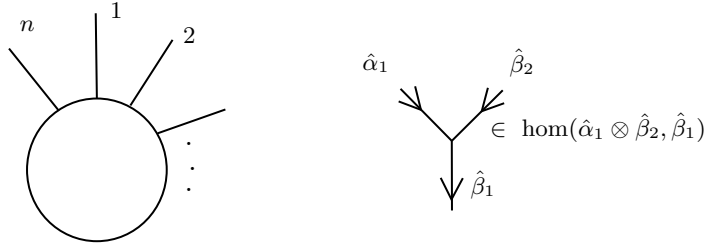
\begin{figure}
    \centering

\tikzset{every picture/.style={line width=0.75pt}}

\begin{tikzpicture}[x=0.75pt,y=0.75pt,yscale=-1,xscale=1]

\draw    (275.86,86.89) -- (301.45,113.07) ;

\draw    (301.45,113.07) -- (326.89,87.62) ;

\draw    (301.45,113.07) -- (301.54,148.08) ;
\draw   (283.15,85.62) -- (286.05,97.45) -- (274.93,93.22) ;
\draw   (327.59,94.12) -- (316.4,98.16) -- (319.48,86.38) ;
\draw   (306.97,131.82) -- (301.3,142.54) -- (295.91,131.67) ;

\draw   (89.42,127.65) .. controls (89.42,107.83) and (105.03,91.76) .. (124.29,91.76) .. controls (143.55,91.76) and (159.16,107.83) .. (159.16,127.65) .. controls (159.16,147.47) and (143.55,163.53) .. (124.29,163.53) .. controls (105.03,163.53) and (89.42,147.47) .. (89.42,127.65) -- cycle ;

\draw    (124.29,91.76) -- (123.72,48.88) ;

\draw    (141.02,95.26) -- (162.08,62.26) ;

\draw    (154.62,109.27) -- (188.44,97.27) ;

\draw    (105.3,97.89) -- (80.43,66.63) ;

\draw (309.28,126.64) node [anchor=north west][inner sep=0.75pt]  [font=\footnotesize,rotate=-359.84]  {$\hat{\beta }_{1}$};

\draw (255.39,67.53) node [anchor=north west][inner sep=0.75pt]  [font=\footnotesize,rotate=-359.84]  {$\hat{\alpha }_{1}$};

\draw (328.68,65.32) node [anchor=north west][inner sep=0.75pt]  [font=\footnotesize,rotate=-359.84]  {$\hat{\beta }_{2}$};

\draw (319.12,97.73) node [anchor=north west][inner sep=0.75pt]  [font=\footnotesize,rotate=-359.84]  {$\in \ \hom(\hat{\alpha }_{1} \otimes \hat{\beta }_{2} ,\hat{\beta }_{1})$};

\draw (129.35,42.21) node [anchor=north west][inner sep=0.75pt]  [font=\footnotesize]  {$1$};

\draw (165.92,54.47) node [anchor=north west][inner sep=0.75pt]  [font=\footnotesize]  {$2$};

\draw (84.27,50.97) node [anchor=north west][inner sep=0.75pt]  [font=\footnotesize]  {$n$};

\draw (166.77,112.23) node [anchor=north west][inner sep=0.75pt]    {$.$};

\draw (168.47,124.49) node [anchor=north west][inner sep=0.75pt]    {$.$};

\draw (166.77,134.99) node [anchor=north west][inner sep=0.75pt]    {$.$};

\end{tikzpicture}

\caption{n-line junction.}
\label{fig: n-line junction}    
\end{figure}

\begin{figure}
    \centering
    
\tikzset{every picture/.style={line width=0.75pt}}
\begin{tikzpicture}[x=0.75pt,y=0.75pt,yscale=-1,xscale=1]

\draw   (154.1,141.73) .. controls (154.1,121.18) and (170.29,104.53) .. (190.26,104.53) .. controls (210.24,104.53) and (226.43,121.18) .. (226.43,141.73) .. controls (226.43,162.27) and (210.24,178.93) .. (190.26,178.93) .. controls (170.29,178.93) and (154.1,162.27) .. (154.1,141.73) -- cycle ;

\draw    (190.26,104.53) -- (189.68,60.07) ;

\draw    (190.85,223.38) -- (190.26,178.93) ;

\draw    (226.43,141.73) -- (267.3,141.73) ;

\draw    (154.1,141.73) -- (110.29,141.73) ;
\draw   (194.97,78.22) -- (190.09,89.19) -- (185,78.32) ;
\draw   (195.85,194.35) -- (190.53,207.13) -- (184.98,194.46) ;
\draw   (244.11,136.39) -- (256.64,141.6) -- (244.43,147.57) ;
\draw   (122.39,136.39) -- (134.92,141.6) -- (122.71,147.57) ;
\draw   (214.77,123.03) -- (209.94,110.05) -- (222.5,115.17) ;
\draw   (207.86,168.94) -- (220.28,163.45) -- (215.81,176.57) ;
\draw   (164.59,160.63) -- (170.49,173.13) -- (157.55,169.15) ;
\draw   (173.04,113.29) -- (160.92,119.45) -- (164.71,106.11) ;

\draw   (361.38,153.52) .. controls (361.38,132.98) and (377.57,116.32) .. (397.54,116.32) .. controls (417.51,116.32) and (433.71,132.98) .. (433.71,153.52) .. controls (433.71,174.07) and (417.51,190.72) .. (397.54,190.72) .. controls (377.57,190.72) and (361.38,174.07) .. (361.38,153.52) -- cycle ;

\draw    (397.54,116.32) -- (396.95,71.87) ;

\draw    (450.13,214.4) -- (421,181.56) ;

\draw    (361.38,153.52) -- (317.57,153.52) ;
\draw   (402.25,90.01) -- (397.36,100.98) -- (392.28,90.11) ;
\draw   (434.25,188.37) -- (438.52,201.56) -- (426.19,195.87) ;
\draw   (329.67,148.19) -- (342.19,153.4) -- (329.99,159.36) ;
\draw   (422.05,134.82) -- (417.22,121.84) -- (429.78,126.97) ;
\draw   (371.87,172.42) -- (377.77,184.93) -- (364.82,180.94) ;
\draw   (380.39,126.74) -- (367.8,131.77) -- (372.71,118.82) ;

\draw (195.32,52.15) node [anchor=north west][inner sep=0.75pt]  [font=\footnotesize]  {$\hat{\alpha }_{1}$};

\draw (194.44,206.39) node [anchor=north west][inner sep=0.75pt]  [font=\footnotesize]  {$\hat{\alpha }_{3}$};

\draw (263.23,146.5) node [anchor=north west][inner sep=0.75pt]  [font=\footnotesize]  {$\hat{\alpha }_{2}$};

\draw (109.76,149.23) node [anchor=north west][inner sep=0.75pt]  [font=\footnotesize]  {$\hat{\alpha }_{4}$};

\draw (221.9,90.07) node [anchor=north west][inner sep=0.75pt]  [font=\footnotesize]  {$\hat{\beta }_{1}$};

\draw (221.02,170.82) node [anchor=north west][inner sep=0.75pt]  [font=\footnotesize]  {$\hat{\beta }_{2}$};

\draw (153.98,172.63) node [anchor=north west][inner sep=0.75pt]  [font=\footnotesize]  {$\hat{\beta }_{3}$};

\draw (146.92,85.53) node [anchor=north west][inner sep=0.75pt]  [font=\footnotesize]  {$\hat{\beta }_{4}$};

\draw (402.6,63.94) node [anchor=north west][inner sep=0.75pt]  [font=\footnotesize]  {$\hat{\alpha }_{1}$};

\draw (448.95,185.69) node [anchor=north west][inner sep=0.75pt]  [font=\footnotesize,rotate=-358.44]  {$\hat{\alpha }_{2}$};

\draw (319.57,156.92) node [anchor=north west][inner sep=0.75pt]  [font=\footnotesize]  {$\hat{\alpha }_{3}$};

\draw (429.17,101.86) node [anchor=north west][inner sep=0.75pt]  [font=\footnotesize]  {$\hat{\beta }_{1}$};

\draw (361.26,184.43) node [anchor=north west][inner sep=0.75pt]  [font=\footnotesize]  {$\hat{\beta }_{2}$};

\draw (354.2,97.33) node [anchor=north west][inner sep=0.75pt]  [font=\footnotesize]  {$\hat{\beta }_{3}$};

\end{tikzpicture}
 \caption{From left to right: 4-line junction and 3-line junction.}
    \label{fig: 4,3-line junctions}
\end{figure}
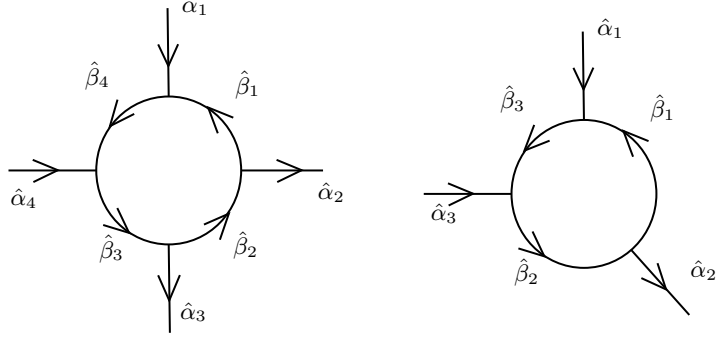

\subsection{Elementary defects}
\label{section: elementary defects}

Our proposed set of defects is built starting from n-line junctions (see left hand side of figure \ref{fig: n-line junction}) defined as follows. We assign orientations and representations to the lines, getting a term like (\ref{eq: unitarity}) for each. In the following we use $\hat{\alpha}_{i}$ for lines associated to $U$s and $\hat{\beta}_{i}$ for lines associated to $V$s: notice that the data associated to this object are representations, the underlying holonomies are integrated over in the path integral. If we zoom into each triple intersection we get the right hand side of figure \ref{fig: n-line junction}. We can then use Clebsch-Gordan coefficients (Appendix \ref{appendix: clebsch-gordan}) to contract the matrix elements coming from the lines.
Here we showcase the definition of the 3 and 4-line junction (see figure \ref{fig: 4,3-line junctions}) that we use in the next sections.

\begin{equation}
\begin{split}
    \mathbb{W}_{3}(\left \{ \hat{\alpha}_{i}, \hat{\beta}_{i}, \nu_{i} \right \}_{i=1}^{3})^{ac}_{b'} :=& R_{\hat{\alpha}_{1}}(U_{1})^{a}_{a'} R_{\hat{\alpha}_{2}}(U_{2})^{b}_{b'} R_{\hat{\alpha}_{3}}(U_{3})^{c}_{c'} R_{\hat{\beta}_{1}}(V_{1}^{-1})^{d}_{d'} R_{\hat{\beta}_{2}}(V_{2}^{-1})^{e}_{e'} R_{\hat{\beta}_{3}}(V_{3}^{-1})^{f}_{f'} \times \\
    & \times C_{\nu_{1}}(\hat{\beta}_{1}, \hat{\alpha}_{1}, \overline{\hat{\beta}}_{3})^{d'a'}_{f} \, C_{\nu_{2}}(\hat{\beta}_{3}, \hat{\alpha}_{3}, \overline{\hat{\beta}}_{2})^{f'c'}_{e} \, \overline{C}_{\nu_{3}}(\hat{\beta}_{1}, \hat{\alpha}_{2}, \overline{\hat{\beta}}_{2})^{e'}_{db} \, .
\end{split}    
\label{eq: W3 definition}
\end{equation}

\begin{equation}
    \begin{split}
    \mathbb{W}_{4}&(\left \{ \hat{\alpha}_{i}, \hat{\beta}_{i}, \nu_{i}\right \}_{i=1}^{4} )^{ad}_{b'c'}  := R_{\hat{\alpha}_{1}}(U_{1})^{a}_{a'} R_{\hat{\alpha}_{2}}(U_{2})^{b}_{b'} R_{\hat{\alpha}_{3}}(U_{3})^{c}_{c'} R_{\hat{\alpha}_{4}}(U_{4})^{d}_{d'} R_{\hat{\beta}_{1}}(V_{1}^{-1})^{e}_{e'} R_{\hat{\beta}_{2}}(V_{2}^{-1})^{f}_{f'} \times \\
    & \times R_{\hat{\beta}_{3}}(V_{3}^{-1})^{g}_{g'} R_{\hat{\beta}_{4}}(V_{4}^{-1})^{h}_{h'} C_{\nu_{1}}(\hat{\beta}_{1}, \hat{\alpha}_{1}, \overline{\hat{\beta}}_{4})^{e'a'}_{h} \, \overline{C}_{\nu_{2}}(\hat{\beta}_{1}, \hat{\alpha}_{2}, \overline{\hat{\beta}}_{2})_{e b}^{f'} \, \times \\
    & \times \overline{C}_{\nu_{3}}(\hat{\beta}_{2}, \hat{\alpha}_{3}, \overline{\hat{\beta}}_{3})_{f c}^{g'} \, C_{\nu_{4}}(\hat{\beta}_{4}, \hat{\alpha}_{4}, \overline{\hat{\beta}}_{3})^{h'd'}_{g} \, .
\end{split}
\label{eq: W4 definition}
\end{equation}
These junctions inserted at a point in the lattice are operators depending on $U$s and $V$s, the dependence on which is implicit on the lhs. 
The junctions depend on some data that we get to choose: the representations associated to each line and the $\nu_{i}$ indices of the Clebsch-Gordan coefficients. The $U$ and $V$ holonomies are degrees of freedom and are integrated over in the path integral. 
We have chosen a convention for the orientation of the $V$ holonomies in such a way that they appear as $V^{-1}$ in the definition of the junction. If we inverted the $\hat{\beta}$ arrows in figure \ref{fig: 4,3-line junctions}, they would appear as $V$ and nothing would change in the computations. \par
The junctions obtained in this way are not gauge invariant, as one can tell by their index structure. In order to build the full gauge invariant defect, we have to compose enough junctions until there are no external legs and hence all indices are contracted. We compose two n-line junctions along a line by identifying the representations $\hat{\alpha}_{i}$ and choosing the orientation consistently. We denote this composition as $\mathbb{W}_{n_{1}} \times_{\hat{\alpha}_{i}} \mathbb{W}_{n_{2}}$. We will later also define a fusion operation given by pushing two junctions together along a line $U_{i}$, denoted as $\mathbb{W}_{n_{1}} \circ_{U_{i}} \mathbb{W}_{n_{2}}$. Fusion is a composition along $\hat{\alpha}_{i}$ followed by squeezing the holonomy $U_i$ to 1. \par
Composition and fusion are operations on the $U$ lines only. We can also apply degenerations: making group elements invisible by setting the associated representation to be trivial. Degeneration can be applied to both $U$s and $V$s. In sections \ref{section: degeneration} and \ref{section: fusion}, we use combinations of these operations to show that the fundamental building block is simply $\mathbb{W}_{3}$. \par

Our basic set of gauge invariant operators is built by composing any number of n-line junctions so that there are no external legs. The defects thus obtained depend on data given by the representations and the $\nu_{i}$ indices. A generic defect is a weighted linear combination of these junctions, and any admissible choice of weight defines a gauge invariant operator in this set.

By "weights" we mean functions defined on the space of admissible representations, summed over all the relevant representations. These functions can be interpreted as projections and will be denoted with the letter $p$. To construct the most general gauge invariant network of defects we have to insert an n-line junction between plaquettes (see figure \ref{fig: gluing of a 3 and a 4-line junction} for examples). This involves integrating out all degrees of freedom with appropriate weights, which include the above projections as well as the weights for $B$ discussed in section \ref{section: the proposal}.  

In the computations of sections \ref{section: degeneration}, \ref{section: fusion} and \ref{section: IRF crossing} we set $B=1$ for simplicity and omit the tracking of theta angles. The algorithm outlined above makes it clear how to reinstate them.

\paragraph{Selection rules}
Even though the defects built so far are gauge invariant, not all of them will be non-trivial due to the symmetries of the theory. Indeed, we can have selection rules corresponding to the generalized symmetries under which the defects are charged. We demonstrate this in two simple examples. In the following examples we omit the $V$s for conciseness: they are not needed as the examples involve no Wilson points.
It is easy to check that everything works exactly the same if we take into account the $V$s as well.

\paragraph{Example 1} Consider the $PSU(N)$ partition function on a 2-sphere with a non-self-intersecting Wilson loop $\hat{R}$ and two different theta angles $\hat{\theta}, \hat{\theta}'$ in the two regions separated by the loop. The relevant partition function is given by

\begin{equation}
    \begin{split}
        Z_{\mathbb{S}^{2}} (\epsilon_{1}, \epsilon_{2}; \hat{\theta}, \hat{\theta'}, \hat{R}) =& \int dU \sum_{\hat{\rho}, \hat{\sigma} \in \widehat{\tilde{G}}} \delta_{\hat{h}_{\hat{\rho}}, \hat{\theta}} \, \delta_{\hat{h}_{\hat{\sigma}}, \hat{\theta}'} \, \dim(\hat{\rho}) \dim(\hat{\sigma}) \chi_{\hat{\sigma}}(U^{-1}) \chi_{\hat{\rho}} (U) \chi_{\hat{R}}(U) \times \\
        & \times e^{-\frac{\epsilon_{1}}{2} c_{2}(\hat{\rho})} e^{-\frac{\epsilon_{2}}{2} c_{2}(\hat{\sigma})} \\
        =& \sum_{\hat{\rho}, \hat{\sigma} \in \widehat{\tilde{G}}} \delta_{\hat{h}_{\hat{\rho}}, \hat{\theta}} \, \delta_{\hat{h}_{\hat{\sigma}}, \hat{\theta}'} \, N^{\hat{\sigma}}_{\hat{\rho} \hat{R}} \dim(\hat{\rho}) \dim(\hat{\sigma}) e^{-\frac{\epsilon_{1}}{2} c_{2}(\hat{\rho})} e^{-\frac{\epsilon_{2}}{2} c_{2}(\hat{\sigma})} \, .
    \end{split}
\end{equation}
The appearance of the fusion numbers $N^{\hat{\sigma}}_{\hat{\rho} \hat{R}}$ yields a selection rule for the Wilson loop: the partition function is non-zero only if the N-ality of $\hat{R}$ is given by the difference of the N-alities of the theta angles (mod $N$), see Appendix \ref{appendix: clebsch-gordan}. \par

\paragraph{Example 2} Consider instead the partition function on a 2-sphere with a Wilson loop $\hat{R}$ and a theta angle $\hat{\theta}$ on one hemisphere, but now we put the $\tilde{G}$ theory on the other hemisphere by setting $B=1$

\begin{equation}
\begin{split}
    Z_{\mathbb{S}^{2}}(\epsilon_{1}, \epsilon_{2}; \hat{\theta}, \hat{R}) =& \frac{1}{|H|} \int dU \sum_{B_{2} \in H} \sum_{\hat{\rho}, \hat{\sigma} \in \widehat{\tilde{G}}} \dim(\hat{\rho}) \dim(\hat{\sigma}) \chi_{\hat{h}_{\hat{\rho}}}(B_{2}) \chi_{\hat{\theta}}(B_{2}^{-1}) e^{-\frac{\epsilon_{1}}{2} c_{2}(\hat{\rho})} e^{-\frac{\epsilon_{2}}{2} c_{2}(\hat{\sigma})} \times \\
    & \times \chi_{\hat{\sigma}}(U^{-1}) \chi_{\hat{\rho}} (U) \chi_{\hat{R}}(U)\\
    =& \sum_{\hat{\sigma}, \hat{\rho} \in \widehat{\tilde{G}}} \dim(\hat{\sigma}) \dim(\hat{\rho}) \, N_{\hat{\rho} \hat{R}}^{\hat{\sigma}} \, \delta_{\hat{h}_{\hat{\rho}}, \hat{\theta}} e^{-\frac{\epsilon_{1}}{2} c_{2}(\hat{\rho})} e^{-\frac{\epsilon_{2}}{2} c_{2}(\hat{\sigma})} \, .
\end{split}
\label{eq: half gauging}
\end{equation}
Here, the Wilson loop behaves as an interface between an $\tilde{G}$ and an $\tilde{G}/H$ theory that selects on the $\tilde{G}$ side the subset of representations with $N$-ality given by the difference between $\hat{\theta}$ and $\hat{R}$. Note that our formalism enables us to control gauging over single plaquettes. This is reminiscent of the domain wall constructions discussed in higher dimensions in e.g. \cite{Gaiotto:2008sd, Gaiotto:2008sa}. \par

\subsection{Degeneration}
\label{section: degeneration}
We now show that 2 and 1-line junctions are degenerate cases of the 3-line junction. We will also explicitly recover a pure Wilson line and a pure Wilson point as degenerate cases of the 2-line junction.

\paragraph{Recovering 2 and 1-line junctions} Starting with the 3-line junction (\ref{eq: W3 definition}) and erasing one line by considering the weight $\delta_{\hat{\alpha}_{3}, \hat{e}}$, or more precisely the projection  $p = \sum_{\hat{\alpha}_{3}} \delta_{\hat{\alpha}_{3}, \hat{e}}$, we obtain the 2-line junction

\begin{equation}
        \begin{split}
    \sum_{\hat{\alpha}_{3}}\delta_{\hat{\alpha}_{3}, \hat{e}} \mathbb{W}_{3}\left (\left \{ \hat{\alpha}_{i}, \hat{\beta}_{i}, \nu_{i} \right \}_{i=1}^{3}\right )^{ac}_{b'} =& R_{\hat{\alpha}_{1}}\left (U_{1}\right )_{a'}^{a} R_{\hat{\alpha}_{2}}\left (U_{2}\right )^{b}_{b'} R_{\hat{e}}\left (U_{3}\right )_{c'}^{c} R_{\hat{\beta}_{1}}\left (V_{1}^{-1}\right )^{d}_{d'} R_{\hat{\beta}_{2}}\left (V_{2}^{-1}\right )^{e}_{e'} \times \\
    & \times  R_{\hat{\beta}_{3}}\left (V_{3}^{-1}\right )^{f}_{f'} C_{\nu_{1}}\left (\hat{\beta}_{1}, \hat{\alpha}_{1}, \overline{\hat{\beta}}_{3}\right )^{d'a'}_{f} \, C_{\nu_{2}}\left (\hat{\beta}_{3}, \hat{e}, \overline{\hat{\beta}}_{2}\right )^{f'c'}_{e} \, \times \\
    & \times \overline{C}_{\nu_{3}}\left (\hat{\beta}_{1}, \hat{\alpha}_{2}, \overline{\hat{\beta}}_{2}\right )^{e'}_{db} \\
    =& R_{\hat{\alpha}_{1}}\left (U_{1}\right )_{a'}^{a} R_{\hat{\alpha}_{2}}\left (U_{2}\right )^{b}_{b'} R_{\hat{\beta}_{1}}\left (V_{1}^{-1}\right )^{d}_{d'} R_{\hat{\beta}_{2}}\left (V_{3}^{-1}V_{2}^{-1}\right )^{f}_{e'} \times \\
    & \times C_{\nu_{1}}\left (\hat{\beta}_{1}, \hat{\alpha}_{1}, \overline{\hat{\beta}}_{2}\right )^{d'a'}_{f}\, \overline{C}_{\nu_{2}}\left (\hat{\beta}_{1}, \hat{\alpha}_{2}, \overline{\hat{\beta}}_{2}\right )^{e'}_{db} \\
    =:& \mathbb{W}_{2}\left (\left \{ \hat{\alpha}_{i}, \hat{\beta}_{i}, \nu_{i} \right \}_{i=1}^{2}\right )^{a}_{b'} \, .
\end{split}    
\label{eq: 3 to 2-line junction}
\end{equation}
Where we used (\ref{eq: special case for CG}) and in the last passage we renamed $\nu_{3}$ to $\nu_{2}$ for convenience. Notice that after the projection $c$ is an index in the trivial representation, therefore it takes a unique value and can be omitted in the final result. In the same way we can erase one more line and get the 1-line junction

\begin{equation}
    \begin{split}
        \sum_{\hat{\alpha}_{2}}\delta_{\hat{\alpha}_{2}, \hat{e}} \mathbb{W}_{2}\left (\left \{ \hat{\alpha}_{i}, \hat{\beta}_{i}, \nu_{i} \right \}_{i=1}^{2}\right )^{a}_{b'} =& R_{\hat{\alpha}_{1}}\left (U_{2}\right )_{a'}^{a} R_{\hat{\beta}_{1}}\left (V_{3}^{-1}V_{2}^{-1}V_{1}^{-1}\right )^{f}_{d'} C_{\nu_{1}}\left (\hat{\beta}_{1}, \hat{\alpha}_{2}, \overline{\hat{\beta}}_{1}\right )^{d'a'}_{f} \\
    =:& \mathbb{W}_{1}\left (\left \{\hat{\alpha}_{1},\hat{\beta}_{1}, \nu_{1}\right \}\right )^{a} \, ,
    \end{split}
\end{equation}

\noindent where again we omit the index in the trivial representation.

\paragraph{Recovering the Wilson line}

\begin{figure}
    \centering
    \tikzset{every picture/.style={line width=0.75pt}}
\begin{tikzpicture}[x=0.75pt,y=0.75pt,yscale=-1,xscale=1]

\draw   (126.47,121.96) .. controls (126.47,102.95) and (142.43,87.54) .. (162.12,87.54) .. controls (181.81,87.54) and (197.77,102.95) .. (197.77,121.96) .. controls (197.77,140.97) and (181.81,156.38) .. (162.12,156.38) .. controls (142.43,156.38) and (126.47,140.97) .. (126.47,121.96) -- cycle ;

\draw [color={rgb, 255:red, 74; green, 144; blue, 226 }  ,draw opacity=1 ]   (162.12,87.54) -- (161.54,46.4) ;

\draw [color={rgb, 255:red, 74; green, 144; blue, 226 }  ,draw opacity=1 ]   (162.7,197.52) -- (162.12,156.38) ;
\draw  [color={rgb, 255:red, 74; green, 144; blue, 226 }  ,draw opacity=1 ] (166.76,63.19) -- (161.94,73.34) -- (156.93,63.28) ;
\draw  [color={rgb, 255:red, 74; green, 144; blue, 226 }  ,draw opacity=1 ] (167.63,170.66) -- (162.39,182.48) -- (156.91,170.75) ;
\draw  [color={rgb, 255:red, 74; green, 144; blue, 226 }  ,draw opacity=1 ] (136.81,139.45) -- (142.63,151.02) -- (129.87,147.33) ;
\draw  [color={rgb, 255:red, 74; green, 144; blue, 226 }  ,draw opacity=1 ] (145.14,95.65) -- (133.2,101.34) -- (136.93,89) ;

\draw  [draw opacity=0] (126.88,120.7) .. controls (127.33,102.31) and (142.73,87.53) .. (161.65,87.53) .. controls (161.8,87.53) and (161.95,87.53) .. (162.11,87.54) -- (161.65,121.54) -- cycle ; \draw  [color={rgb, 255:red, 74; green, 144; blue, 226 }  ,draw opacity=1 ] (126.88,120.7) .. controls (127.33,102.31) and (142.73,87.53) .. (161.65,87.53) .. controls (161.8,87.53) and (161.95,87.53) .. (162.11,87.54) ;  

\draw  [draw opacity=0] (161.83,156.39) .. controls (142.75,156.4) and (127.07,141.87) .. (126.63,123.58) .. controls (126.63,123.43) and (126.63,123.27) .. (126.63,123.11) -- (161.84,122.8) -- cycle ; \draw  [color={rgb, 255:red, 74; green, 144; blue, 226 }  ,draw opacity=1 ] (161.83,156.39) .. controls (142.75,156.4) and (127.07,141.87) .. (126.63,123.58) .. controls (126.63,123.43) and (126.63,123.27) .. (126.63,123.11) ;  

\draw    (255.52,120.52) -- (295.26,121.32) ;
\draw [shift={(297.26,121.36)}, rotate = 181.15] [color={rgb, 255:red, 0; green, 0; blue, 0 }  ][line width=0.75]    (10.93,-3.29) .. controls (6.95,-1.4) and (3.31,-0.3) .. (0,0) .. controls (3.31,0.3) and (6.95,1.4) .. (10.93,3.29)   ;

\draw [color={rgb, 255:red, 74; green, 144; blue, 226 }  ,draw opacity=1 ]   (346.59,45.8) -- (347.81,196.12) ;
\draw [shift={(347.83,198.12)}, rotate = 269.53] [color={rgb, 255:red, 74; green, 144; blue, 226 }  ,draw opacity=1 ][line width=0.75]    (10.93,-3.29) .. controls (6.95,-1.4) and (3.31,-0.3) .. (0,0) .. controls (3.31,0.3) and (6.95,1.4) .. (10.93,3.29)   ;
\draw  [color={rgb, 255:red, 208; green, 2; blue, 27 }  ,draw opacity=1 ] (206.12,129.36) -- (188.29,111.57)(187.99,129.7) -- (206.42,111.24) ;

\draw (166.97,38.23) node [anchor=north west][inner sep=0.75pt]  [font=\footnotesize]  {$\hat{\alpha }_{1}$};

\draw (169.41,183.35) node [anchor=north west][inner sep=0.75pt]  [font=\footnotesize]  {$\hat{\alpha }_{2}$};

\draw (210.18,110.18) node [anchor=north west][inner sep=0.75pt]  [font=\footnotesize]  {$\hat{\beta }_{1}$};

\draw (106.31,110.9) node [anchor=north west][inner sep=0.75pt]  [font=\footnotesize]  {$\hat{\beta }_{2}$};

\draw (353.5,48.31) node [anchor=north west][inner sep=0.75pt]  [font=\footnotesize]  {$\hat{\alpha }_{1}$};

\end{tikzpicture}

    \caption{Moves to get a Wilson line from a 2-line junction.}
    \label{fig: Getting Wilson line from 2-line}
\end{figure}
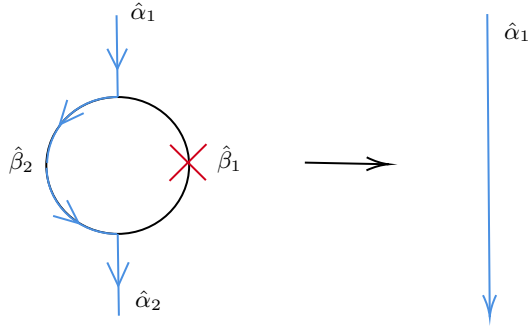

Starting with the 2-line junction as defined in (\ref{eq: 3 to 2-line junction}), we can perform the moves depicted in figure \ref{fig: Getting Wilson line from 2-line} to get the local version of a Wilson line. This corresponds to the following projection

\begin{equation}
    \sum_{\hat{\alpha}_{2}, \hat{\beta}_{1}, \hat{\beta}_{2}} \delta_{\hat{\beta}_{1}, \hat{e}} \mathbb{W}_{2} \left (\left \{ \hat{\alpha}_{i}, \hat{\beta}_{i}, \nu_{i} \right \}_{i=1}^{2}\right )^{a}_{b'} = R_{\hat{\alpha}_{1}}\left (U_{1}V_{1}U_{2}\right )^{a}_{b'}
    =: \mathbb{W}_{|} \left( \left \{ \hat{\alpha}_{1} \right\} \right )^{a}_{b'} \, .
\end{equation}
We used (\ref{eq: special case for CG}) and we summed over all the extra representations.

\paragraph{Recovering the Wilson point} 

\begin{figure}
    \centering
    \tikzset{every picture/.style={line width=0.75pt}}
\begin{tikzpicture}[x=0.75pt,y=0.75pt,yscale=-1,xscale=1]

\draw   (144.72,159.58) .. controls (144.72,139.75) and (160.77,123.67) .. (180.57,123.67) .. controls (200.37,123.67) and (216.42,139.75) .. (216.42,159.58) .. controls (216.42,179.41) and (200.37,195.49) .. (180.57,195.49) .. controls (160.77,195.49) and (144.72,179.41) .. (144.72,159.58) -- cycle ;

\draw [color={rgb, 255:red, 0; green, 0; blue, 0 }  ,draw opacity=1 ]   (180.57,123.67) -- (179.98,80.76) ;

\draw [color={rgb, 255:red, 0; green, 0; blue, 0 }  ,draw opacity=1 ]   (181.15,238.4) -- (180.57,195.49) ;
\draw  [color={rgb, 255:red, 0; green, 0; blue, 0 }  ,draw opacity=1 ] (150.95,151.37) -- (145.28,163.51) -- (140.17,151.1) ;
\draw  [color={rgb, 255:red, 208; green, 2; blue, 27 }  ,draw opacity=1 ] (171.37,112.02) -- (189.77,93.29)(171.22,93.44) -- (189.92,111.87) ;

\draw    (294.61,158.08) -- (334.58,158.91) ;
\draw [shift={(336.58,158.95)}, rotate = 181.2] [color={rgb, 255:red, 0; green, 0; blue, 0 }  ][line width=0.75]    (10.93,-3.29) .. controls (6.95,-1.4) and (3.31,-0.3) .. (0,0) .. controls (3.31,0.3) and (6.95,1.4) .. (10.93,3.29)   ;
\draw  [color={rgb, 255:red, 208; green, 2; blue, 27 }  ,draw opacity=1 ] (171.37,220.62) -- (189.77,201.89)(171.22,202.04) -- (189.92,220.47) ;
\draw  [color={rgb, 255:red, 0; green, 0; blue, 0 }  ,draw opacity=1 ] (210.7,168.22) -- (215.92,155.86) -- (221.48,168.05) ;
\draw   (374.03,147.52) .. controls (379.23,142.47) and (387.54,142.6) .. (392.57,147.81) .. controls (397.61,153.03) and (397.48,161.34) .. (392.28,166.39) .. controls (387.07,171.44) and (378.77,171.3) .. (373.73,166.09) .. controls (368.69,160.88) and (368.83,152.56) .. (374.03,147.52) -- cycle ; \draw   (374.03,147.52) -- (392.28,166.39) ; \draw   (392.57,147.81) -- (373.73,166.09) ;

\draw (185.49,72.72) node [anchor=north west][inner sep=0.75pt]  [font=\footnotesize]  {$\hat{\alpha }_{1}$};

\draw (191.82,221.81) node [anchor=north west][inner sep=0.75pt]  [font=\footnotesize]  {$\hat{\alpha }_{2}$};

\draw (224.85,149.26) node [anchor=north west][inner sep=0.75pt]  [font=\footnotesize]  {$\hat{\beta }_{1}$};

\draw (122.98,146.21) node [anchor=north west][inner sep=0.75pt]  [font=\footnotesize]  {$\hat{\beta }_{2}$};

\draw (394.33,120.64) node [anchor=north west][inner sep=0.75pt]  [font=\footnotesize]  {$\hat{\beta }_{1}$};

\end{tikzpicture}

    \caption{Moves to get a Wilson point from a 2-line junction.}
    \label{fig: getting Wilson point from 2-line}
\end{figure}
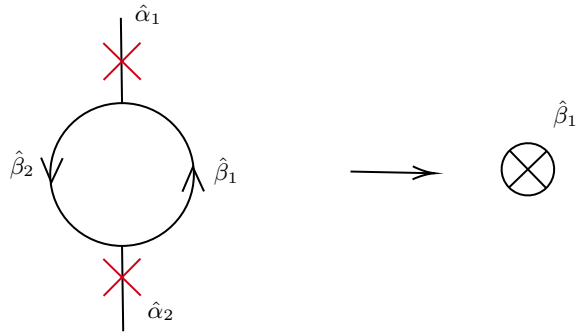

Starting with a 2-line junction we can recover the Wilson point by considering a projection corresponding to the moves depicted in figure \ref{fig: getting Wilson point from 2-line}

\begin{equation}
    \sum_{\hat{\alpha}_{1}, \hat{\alpha}_{2}, \hat{\beta}_{2}} \delta_{\hat{\alpha}_{1}, \hat{e}} \delta_{\hat{\alpha}_{2}, \hat{e}} \mathbb{W}_{2}\left (\left \{ \hat{\alpha}_{i}, \hat{\beta}_{i}, \nu_{i} \right \}_{i=1}^{2}\right )^{a}_{b'} = \chi_{\hat{\beta}_{1}}\left (V_{1}^{-1}V_{2}^{-1}\right ) =: \mathbb{W}_{\bullet} \left(  \hat{\beta}_{1} \right) \, .
\end{equation}
We used (\ref{eq: special case for CG}) and omitted the indices in the trivial representation. This is our definition of a Wilson point. \par

\subsection{Fusion}
\label{section: fusion}

We can build any defect by composing n-line junctions.
We showed how to get $\left (n<3\right )$-line junctions by deleting lines. We claim that we can get $(n>3)$-line junctions by fusing 3-line junctions. Specifically, we will now show that we can get 4-line junctions by fusing two $\mathbb{W}_{3}$. We verify this as an expectation value in a complete network rather than an operator equation.

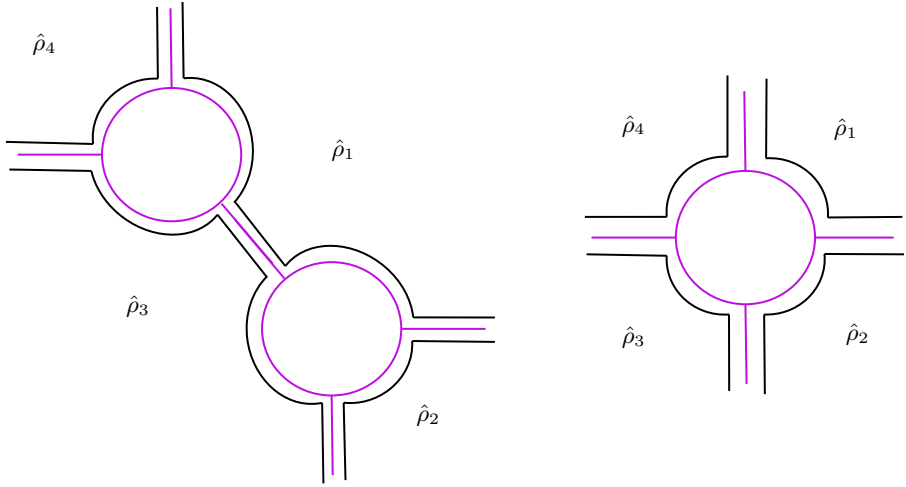
\begin{figure}
    \centering
    \tikzset{every picture/.style={line width=0.75pt}}

\begin{tikzpicture}[x=0.75pt,y=0.75pt,yscale=-1,xscale=1]

\draw  [color={rgb, 255:red, 189; green, 16; blue, 224 }  ,draw opacity=1 ] (431.51,121.63) .. controls (431.51,103.11) and (447.06,88.1) .. (466.25,88.1) .. controls (485.44,88.1) and (501,103.11) .. (501,121.63) .. controls (501,140.16) and (485.44,155.17) .. (466.25,155.17) .. controls (447.06,155.17) and (431.51,140.16) .. (431.51,121.63) -- cycle ;

\draw [color={rgb, 255:red, 189; green, 16; blue, 224 }  ,draw opacity=1 ]   (466.25,88.1) -- (465.69,48.02) ;

\draw [color={rgb, 255:red, 189; green, 16; blue, 224 }  ,draw opacity=1 ]   (466.82,195.25) -- (466.25,155.17) ;

\draw [color={rgb, 255:red, 189; green, 16; blue, 224 }  ,draw opacity=1 ]   (501,121.63) -- (540.26,121.63) ;

\draw [color={rgb, 255:red, 189; green, 16; blue, 224 }  ,draw opacity=1 ]   (431.51,121.63) -- (389.42,121.63) ;

\draw    (457.29,40.07) -- (457.29,80.97) ;

\draw  [draw opacity=0] (426.98,111.18) .. controls (426.23,102.91) and (429.38,94.36) .. (436.28,88.27) .. controls (442.24,83.01) and (449.84,80.59) .. (457.26,80.97) -- (455.34,108.37) -- cycle ; \draw   (426.98,111.18) .. controls (426.23,102.91) and (429.38,94.36) .. (436.28,88.27) .. controls (442.24,83.01) and (449.84,80.59) .. (457.26,80.97) ;  

\draw    (426.98,111.21) -- (386.11,111.23) ;

\draw  [draw opacity=0] (476.51,81.78) .. controls (485.11,81.16) and (493.96,84.34) .. (500.18,91.12) .. controls (505.57,96.98) and (507.94,104.4) .. (507.41,111.59) -- (479.06,109.18) -- cycle ; \draw   (476.51,81.78) .. controls (485.11,81.16) and (493.96,84.34) .. (500.18,91.12) .. controls (505.57,96.98) and (507.94,104.4) .. (507.41,111.59) ;  

\draw  [draw opacity=0] (505.7,129.91) .. controls (506.51,138.17) and (503.41,146.74) .. (496.56,152.88) .. controls (490.63,158.18) and (483.04,160.64) .. (475.62,160.32) -- (477.37,132.91) -- cycle ; \draw   (505.7,129.91) .. controls (506.51,138.17) and (503.41,146.74) .. (496.56,152.88) .. controls (490.63,158.18) and (483.04,160.64) .. (475.62,160.32) ;  

\draw  [draw opacity=0] (458.11,160.29) .. controls (449.51,160.97) and (440.64,157.86) .. (434.36,151.13) .. controls (428.93,145.31) and (426.5,137.91) .. (426.97,130.72) -- (455.34,132.91) -- cycle ; \draw   (458.11,160.29) .. controls (449.51,160.97) and (440.64,157.86) .. (434.36,151.13) .. controls (428.93,145.31) and (426.5,137.91) .. (426.97,130.72) ;  

\draw    (476.57,81.77) -- (476.79,41.71) ;

\draw    (507.41,111.52) -- (544.58,111.23) ;

\draw    (505.7,129.88) -- (545.43,130.05) ;

\draw    (475.59,160.31) -- (475.94,200.39) ;

\draw    (458.05,160.3) -- (458.14,198.76) ;

\draw    (426.97,130.78) -- (386.96,130.05) ;

\draw  [color={rgb, 255:red, 189; green, 16; blue, 224 }  ,draw opacity=1 ] (145.07,79.92) .. controls (145.07,61.39) and (160.63,46.38) .. (179.82,46.38) .. controls (199.01,46.38) and (214.56,61.39) .. (214.56,79.92) .. controls (214.56,98.44) and (199.01,113.45) .. (179.82,113.45) .. controls (160.63,113.45) and (145.07,98.44) .. (145.07,79.92) -- cycle ;

\draw [color={rgb, 255:red, 189; green, 16; blue, 224 }  ,draw opacity=1 ]   (179.82,46.38) -- (179.25,6.3) ;

\draw [color={rgb, 255:red, 189; green, 16; blue, 224 }  ,draw opacity=1 ]   (230.34,134.81) -- (204.76,104.57) ;

\draw [color={rgb, 255:red, 189; green, 16; blue, 224 }  ,draw opacity=1 ]   (145.07,79.92) -- (102.98,79.92) ;

\draw  [color={rgb, 255:red, 189; green, 16; blue, 224 }  ,draw opacity=1 ] (294.45,167.54) .. controls (294.44,186.06) and (278.88,201.07) .. (259.69,201.06) .. controls (240.5,201.05) and (224.95,186.03) .. (224.96,167.51) .. controls (224.96,148.99) and (240.53,133.98) .. (259.71,133.99) .. controls (278.9,133.99) and (294.45,149.01) .. (294.45,167.54) -- cycle ;

\draw [color={rgb, 255:red, 189; green, 16; blue, 224 }  ,draw opacity=1 ]   (259.69,201.06) -- (260.24,241.14) ;

\draw [color={rgb, 255:red, 189; green, 16; blue, 224 }  ,draw opacity=1 ]   (211.54,112.75) -- (236.11,142.2) ;

\draw [color={rgb, 255:red, 189; green, 16; blue, 224 }  ,draw opacity=1 ]   (294.45,167.54) -- (336.54,167.55) ;

\draw  [draw opacity=0] (140.66,74.52) .. controls (139.56,65.74) and (142.95,56.32) .. (150.57,49.59) .. controls (157,43.9) and (165.15,41.29) .. (172.93,41.71) -- (169.29,69.33) -- cycle ; \draw   (140.66,74.52) .. controls (139.56,65.74) and (142.95,56.32) .. (150.57,49.59) .. controls (157,43.9) and (165.15,41.29) .. (172.93,41.71) ;  

\draw  [draw opacity=0] (202.73,110.1) .. controls (202.29,110.66) and (201.82,111.21) .. (201.32,111.74) .. controls (189.91,123.86) and (168.22,122.77) .. (152.88,109.31) .. controls (145.97,103.25) and (141.53,95.72) .. (139.77,88.16) -- (173.55,87.37) -- cycle ; \draw   (202.73,110.1) .. controls (202.29,110.66) and (201.82,111.21) .. (201.32,111.74) .. controls (189.91,123.86) and (168.22,122.77) .. (152.88,109.31) .. controls (145.97,103.25) and (141.53,95.72) .. (139.77,88.16) ;  

\draw  [draw opacity=0] (254.89,204.77) .. controls (254.2,204.91) and (253.5,205.03) .. (252.8,205.11) .. controls (235.96,207.15) and (220.21,192.71) .. (217.63,172.86) .. controls (215.98,160.22) and (220.07,148.43) .. (227.61,141.01) -- (248.12,169.16) -- cycle ; \draw   (254.89,204.77) .. controls (254.2,204.91) and (253.5,205.03) .. (252.8,205.11) .. controls (235.96,207.15) and (220.21,192.71) .. (217.63,172.86) .. controls (215.98,160.22) and (220.07,148.43) .. (227.61,141.01) ;  

\draw  [draw opacity=0] (185.89,41.53) .. controls (186.58,41.44) and (187.29,41.38) .. (188,41.35) .. controls (204.95,40.63) and (219.43,56.26) .. (220.35,76.25) .. controls (220.84,87.16) and (217.19,97.12) .. (211.02,104.03) -- (189.65,77.55) -- cycle ; \draw   (185.89,41.53) .. controls (186.58,41.44) and (187.29,41.38) .. (188,41.35) .. controls (204.95,40.63) and (219.43,56.26) .. (220.35,76.25) .. controls (220.84,87.16) and (217.19,97.12) .. (211.02,104.03) ;  

\draw  [draw opacity=0] (236.58,135.96) .. controls (237.02,135.4) and (237.49,134.86) .. (237.98,134.32) .. controls (249.32,122.14) and (271.02,123.09) .. (286.44,136.46) .. controls (294.67,143.59) and (299.46,152.82) .. (300.41,161.72) -- (265.92,158.53) -- cycle ; \draw   (236.58,135.96) .. controls (237.02,135.4) and (237.49,134.86) .. (237.98,134.32) .. controls (249.32,122.14) and (271.02,123.09) .. (286.44,136.46) .. controls (294.67,143.59) and (299.46,152.82) .. (300.41,161.72) ;  

\draw  [draw opacity=0] (299.88,173.74) .. controls (300.45,182.55) and (296.5,191.75) .. (288.52,198.04) .. controls (281.78,203.35) and (273.51,205.52) .. (265.79,204.69) -- (270.98,177.3) -- cycle ; \draw   (299.88,173.74) .. controls (300.45,182.55) and (296.5,191.75) .. (288.52,198.04) .. controls (281.78,203.35) and (273.51,205.52) .. (265.79,204.69) ;  

\draw    (211.4,103.59) -- (236.59,135.96) ;

\draw    (202.73,110.11) -- (227.24,141.37) ;

\draw    (139.76,88.11) -- (98.83,87.4) ;

\draw    (254.96,204.76) -- (255.6,245.27) ;

\draw    (265.7,204.68) -- (266.62,243.63) ;

\draw    (299.88,173.66) -- (341.2,174.1) ;

\draw    (300.41,161.77) -- (341.2,161.83) ;

\draw    (140.66,74.55) -- (97.13,73.49) ;

\draw    (185.81,41.53) -- (185.27,3.14) ;

\draw    (172.96,41.71) -- (172.56,5.6) ;

\draw (508.95,60.23) node [anchor=north west][inner sep=0.75pt]  [font=\footnotesize]  {$\hat{\rho }_{1}$};

\draw (403.02,58.59) node [anchor=north west][inner sep=0.75pt]  [font=\footnotesize]  {$\hat{\rho }_{4}$};

\draw (403.02,164.93) node [anchor=north west][inner sep=0.75pt]  [font=\footnotesize]  {$\hat{\rho }_{3}$};

\draw (514.88,163.29) node [anchor=north west][inner sep=0.75pt]  [font=\footnotesize]  {$\hat{\rho }_{2}$};

\draw (108.96,17.69) node [anchor=north west][inner sep=0.75pt]  [font=\footnotesize]  {$\hat{\rho }_{4}$};

\draw (258.11,70.86) node [anchor=north west][inner sep=0.75pt]  [font=\footnotesize]  {$\hat{\rho }_{1}$};

\draw (155.57,149.39) node [anchor=north west][inner sep=0.75pt]  [font=\footnotesize]  {$\hat{\rho }_{3}$};

\draw (300.48,204.19) node [anchor=north west][inner sep=0.75pt]  [font=\footnotesize]  {$\hat{\rho }_{2}$};
\end{tikzpicture}

    \caption{Gluing of a 3 and a 4-line junction (in pink) in the middle of four plaquettes $\hat{\rho}_{i}$.}
    \label{fig: gluing of a 3 and a 4-line junction}
\end{figure}

To compute the expectation value of a 4-line junction we glue together four plaquettes with the junction in the middle (see right hand side of figure \ref{fig: gluing of a 3 and a 4-line junction}) and do not specify what is going on far away from this by calling $\mathcal{U}_{i}$ the unspecified part of Migdal's factors. In this way we get the local version of the expectation value and we do not commit to a particular topology. We have

\begin{equation}
    \begin{split}
        \left <\mathbb{W}_{4}\right> =& \int dU_{i} dV_{i} \sum_{\hat{\rho}_{i}} \prod_{i} \dim\left (\hat{\rho}_{i}\right ) \chi_{\hat{\rho}_{1}}\left (U_{1}V_{1}U_{2}\mathcal{U}_{1}\right ) \chi_{\hat{\rho}_{2}}\left (U_{2}^{-1}V_{2}U_{3}\mathcal{U}_{2}\right ) \chi_{\hat{\rho}_{3}}\left (U_{3}^{-1}V_{3}U_{4}^{-1}\mathcal{U}_{3}\right ) \times \\
        & \times \chi_{\hat{\rho}_{4}}\left (U_{4}V_{4}U_{1}^{-1}\mathcal{U}_{4}\right )  \mathbb{W}_{4} \, e^{-\frac{\epsilon_{i}}{2} c_{2}\left (\hat{\rho}_{i}\right )} \\   
    =& \sum_{\mu_{i},\nu_{i}} R_{\hat{\rho}_{1}}\left (\mathcal{U}_{1}\right )^{j'}_{i} \, \epsilon_{\mu_{1}}\left (\hat{\rho}_{1}, \hat{\alpha}_{1}, \overline{\hat{\rho}}_{4}\right )^{i \cdot}_{p'} \, R_{\hat{\rho}_{4}}\left (\mathcal{U}_{4}\right )^{p'}_{o} \, \epsilon_{\mu_{4}}\left (\hat{\rho}_{4}, \hat{\alpha}_{4}, \overline{\hat{\rho}}_{3}\right )^{o \cdot}_{n'} \, R_{\hat{\rho}_{3}}\left (\mathcal{U}_{3}\right )^{n'}_{m}  \, \times \\
    &\times \overline{\epsilon}_{\mu_{3}}\left (\hat{\rho}_{2}, \hat{\alpha}_{3}, \overline{\hat{\rho}}_{3}\right )^{m}_{l' \cdot} \, R_{\hat{\rho}_{2}}\left (\mathcal{U}_{2}\right )^{l'}_{k} \overline{\epsilon}_{\mu_{2}}\left (\hat{\rho}_{1}, \hat{\alpha}_{2}, \overline{\hat{\rho}}_{2}\right )^{k}_{j' \cdot} \, \overline{\epsilon}_{\mu_{1}}\left (\hat{\rho}_{1}, \hat{\alpha}_{1}, \overline{\hat{\rho}}_{4}\right )^{h}_{i' a'} \,  \times \\
    &\times \epsilon_{\mu_{2}}\left (\hat{\rho}_{1}, \hat{\alpha}_{2}, \overline{\hat{\rho}}_{2}\right )^{j b}_{f'} \, \epsilon_{\mu_{3}}\left (\hat{\rho}_{2}, \hat{\alpha}_{3}, \overline{\hat{\rho}}_{3}\right )^{l c}_{m'} \, \overline{\epsilon}_{\mu_{4}}\left (\hat{\rho}_{4}, \hat{\alpha}_{4}, \overline{\hat{\rho}}_{3}\right )^{n}_{o' d'} C_{\nu_{1}}\left (\hat{\rho_{1}}, \hat{\alpha}_{1}, \overline{\hat{\rho}}_{4}\right )^{i'a'}_{h} \, \times \\
    &\times \overline{C}_{\nu_{2}}\left (\hat{\rho}_{1}, \hat{\alpha}_{2}, \overline{\hat{\rho}}_{2}\right )_{j b}^{f'} 
 \, \overline{C}_{\nu_{3}}\left (\hat{\rho}_{2}, \hat{\alpha}_{3}, \overline{\hat{\rho}}_{3}\right )_{lc}^{m'} \, C_{\nu_{4}}\left (\hat{\rho}_{4}, \hat{\alpha}_{4}, \overline{\hat{\rho}}_{3}\right )^{o'd'}_{n} \prod_{i} e^{-\frac{\epsilon_{i}}{2} c_{2}\left (\hat{\rho}_{i}\right )} \,.
    \end{split}
    \label{eq: W4 exp value}
\end{equation}
We have denoted by $\cdot$ the indices that are missing because we did not specify the surroundings of the junction, these correspond to the uncontracted indices in the definition (\ref{eq: W4 definition}). This result can be derived using (\ref{eq: integral for fusion number}) for the $U$s; (\ref{eq: unitarity}) and (\ref{eq: schur orthogonality}) for the $V$s. \par
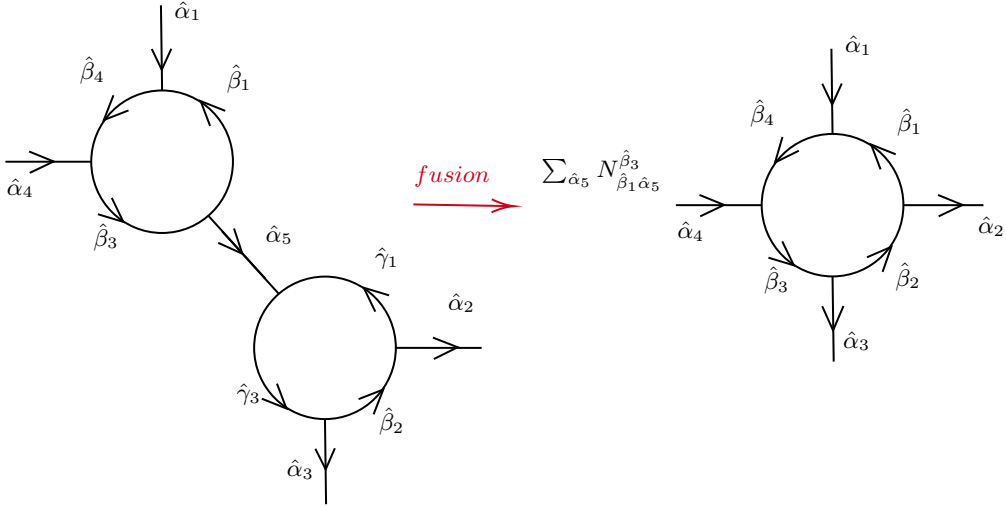
\begin{figure}
    \centering  
\tikzset{every picture/.style={line width=0.75pt}}
\begin{tikzpicture}[x=0.75pt,y=0.75pt,yscale=-1,xscale=1]

\draw   (133.26,83.04) .. controls (133.26,63.25) and (149.09,47.2) .. (168.63,47.2) .. controls (188.16,47.2) and (203.99,63.25) .. (203.99,83.04) .. controls (203.99,102.84) and (188.16,118.88) .. (168.63,118.88) .. controls (149.09,118.88) and (133.26,102.84) .. (133.26,83.04) -- cycle ;

\draw    (168.63,47.2) -- (168.05,4.37) ;

\draw    (220.05,141.7) -- (191.56,110.05) ;

\draw    (133.26,83.04) -- (90.42,83.04) ;
\draw   (173.23,26.22) -- (168.45,36.79) -- (163.48,26.32) ;
\draw   (204.52,116.62) -- (208.69,129.32) -- (196.64,123.84) ;
\draw   (102.25,77.9) -- (114.5,82.92) -- (102.56,88.67) ;
\draw   (192.59,65.03) -- (187.87,52.52) -- (200.15,57.46) ;
\draw   (143.52,101.25) -- (149.29,113.3) -- (136.63,109.46) ;
\draw   (151.85,57.24) -- (139.54,62.08) -- (144.35,49.61) ;

\draw   (285.3,176.67) .. controls (285.29,196.47) and (269.45,212.51) .. (249.92,212.5) .. controls (230.39,212.49) and (214.56,196.44) .. (214.57,176.65) .. controls (214.58,156.86) and (230.41,140.82) .. (249.95,140.82) .. controls (269.48,140.83) and (285.31,156.88) .. (285.3,176.67) -- cycle ;

\draw    (249.92,212.5) -- (250.48,255.33) ;

\draw    (198.53,117.99) -- (227.01,149.64) ;

\draw    (285.3,176.67) -- (328.14,176.69) ;
\draw   (254.88,227.15) -- (250.28,237.8) -- (245.13,227.42) ;
\draw   (303.82,171.51) -- (316.15,176.35) -- (304.3,182.28) ;
\draw   (225.96,194.67) -- (230.68,207.18) -- (218.4,202.23) ;
\draw   (275.05,158.46) -- (269.28,146.41) -- (281.94,150.26) ;
\draw   (266.7,202.47) -- (279.02,197.63) -- (274.2,210.1) ;

\draw   (467.93,104.89) .. controls (467.93,85.1) and (483.76,69.06) .. (503.29,69.06) .. controls (522.83,69.06) and (538.66,85.1) .. (538.66,104.89) .. controls (538.66,124.69) and (522.83,140.73) .. (503.29,140.73) .. controls (483.76,140.73) and (467.93,124.69) .. (467.93,104.89) -- cycle ;

\draw    (503.29,69.06) -- (502.72,26.22) ;

\draw    (503.87,183.57) -- (503.29,140.73) ;

\draw    (538.66,104.89) -- (578.62,104.89) ;

\draw    (467.93,104.89) -- (425.09,104.89) ;
\draw   (507.9,43.71) -- (503.12,54.27) -- (498.15,43.8) ;
\draw   (508.76,155.59) -- (503.56,167.9) -- (498.13,155.7) ;
\draw   (555.95,99.76) -- (568.2,104.78) -- (556.27,110.52) ;
\draw   (436.92,99.76) -- (449.17,104.78) -- (437.23,110.52) ;
\draw   (527.26,86.88) -- (522.54,74.37) -- (534.82,79.31) ;
\draw   (520.5,131.11) -- (532.64,125.83) -- (528.27,138.46) ;
\draw   (478.19,123.11) -- (483.96,135.15) -- (471.3,131.31) ;
\draw   (486.45,77.5) -- (474.6,83.43) -- (478.3,70.57) ;

\draw [color={rgb, 255:red, 208; green, 2; blue, 27 }  ,draw opacity=1 ]   (294.06,104.52) -- (342.09,105.36) ;
\draw [shift={(344.09,105.39)}, rotate = 181] [color={rgb, 255:red, 208; green, 2; blue, 27 }  ,draw opacity=1 ][line width=0.75]    (10.93,-3.29) .. controls (6.95,-1.4) and (3.31,-0.3) .. (0,0) .. controls (3.31,0.3) and (6.95,1.4) .. (10.93,3.29)   ;

\draw (173.36,0.7) node [anchor=north west][inner sep=0.75pt]  [font=\footnotesize]  {$\hat{\alpha }_{1}$};

\draw (218.68,113.63) node [anchor=north west][inner sep=0.75pt]  [font=\footnotesize,rotate=-358.44]  {$\hat{\alpha }_{5}$};

\draw (89.69,89.86) node [anchor=north west][inner sep=0.75pt]  [font=\footnotesize]  {$\hat{\alpha }_{4}$};

\draw (199.37,32.79) node [anchor=north west][inner sep=0.75pt]  [font=\footnotesize]  {$\hat{\beta }_{1}$};

\draw (132.96,112.34) node [anchor=north west][inner sep=0.75pt]  [font=\footnotesize]  {$\hat{\beta }_{3}$};

\draw (126.05,28.42) node [anchor=north west][inner sep=0.75pt]  [font=\footnotesize]  {$\hat{\beta }_{4}$};

\draw (309.66,147.85) node [anchor=north west][inner sep=0.75pt]  [font=\footnotesize,rotate=-358.9]  {$\hat{\alpha }_{2}$};

\draw (275.54,205.06) node [anchor=north west][inner sep=0.75pt]  [font=\footnotesize,rotate=-0.24]  {$\hat{\beta }_{2}$};
 
\draw (273.48,124.83) node [anchor=north west][inner sep=0.75pt]  [font=\footnotesize]  {$\hat{\gamma }_{1}$};
 
\draw (204.48,193.01) node [anchor=north west][inner sep=0.75pt]  [font=\footnotesize]  {$\hat{\gamma }_{3}$};
 
\draw (229.42,231.47) node [anchor=north west][inner sep=0.75pt]  [font=\footnotesize]  {$\hat{\alpha }_{3}$};
 
\draw (508.03,18.18) node [anchor=north west][inner sep=0.75pt]  [font=\footnotesize]  {$\hat{\alpha }_{1}$};
 
\draw (507.16,166.78) node [anchor=north west][inner sep=0.75pt]  [font=\footnotesize]  {$\hat{\alpha }_{3}$};
 
\draw (574.44,109.09) node [anchor=north west][inner sep=0.75pt]  [font=\footnotesize]  {$\hat{\alpha }_{2}$};
 
\draw (424.36,111.71) node [anchor=north west][inner sep=0.75pt]  [font=\footnotesize]  {$\hat{\alpha }_{4}$};
 
\draw (534.04,54.64) node [anchor=north west][inner sep=0.75pt]  [font=\footnotesize]  {$\hat{\beta }_{1}$};
 
\draw (533.18,132.44) node [anchor=north west][inner sep=0.75pt]  [font=\footnotesize]  {$\hat{\beta }_{2}$};
 
\draw (467.62,134.19) node [anchor=north west][inner sep=0.75pt]  [font=\footnotesize]  {$\hat{\beta }_{3}$};
 
\draw (460.72,50.27) node [anchor=north west][inner sep=0.75pt]  [font=\footnotesize]  {$\hat{\beta }_{4}$};
 
\draw (356.58,75.8) node [anchor=north west][inner sep=0.75pt]  [font=\footnotesize]  {$\sum _{\hat{\alpha }_{5}} N_{\hat{\beta }_{1}\hat{\alpha }_{5}}^{\hat{\beta }_{3}}$};
 
\draw (292.98,83.37) node [anchor=north west][inner sep=0.75pt]  [font=\footnotesize,color={rgb, 255:red, 208; green, 2; blue, 27 }  ,opacity=1 ]  {$fusion$};
\end{tikzpicture}
\caption{Fusion of 3-line junctions.}
    \label{fig: fusion of 3-line junctions}
\end{figure}

Using one of the operations we defined in section \ref{section: elementary defects}, we can compose two 3-line junctions along an external line. Based on the way we named things in the left hand side of figure \ref{fig: fusion of 3-line junctions}, we compose along $\hat{\alpha}_{5}$ and get

\begin{equation}
    \begin{split}
        \left (\mathbb{W}_{3} \times_{\hat{\alpha}_{5}} \mathbb{W}_{3} \right)^{\cdot \cdot}_{\cdot \cdot} =& R_{\hat{\alpha}_{1}}\left (U_{1}\right )^{\cdot}_{a'} R_{\hat{\alpha}_{2}}\left (U_{2}\right )_{\cdot}^{b} 
        R_{\hat{\alpha}_{3}}\left (U_{3}\right )_{\cdot}^{c} R_{\hat{\alpha}_{4}}\left (U_{4}\right )^{\cdot}_{d'} R_{\hat{\alpha}_{5}}\left (U_{5}\right )^{j_{3}}_{j'_{3}} \overline{R_{\hat{\beta}_{1}}\left (V_{1}\right )}^{e'}_{e} 
        \overline{R_{\hat{\beta}_{2}}\left (V_{2}\right )}^{f'}_{f} \times \\
        & \times \overline{R_{\hat{\beta}_{3}}\left (V_{3}\right )}^{g'}_{g} \overline{R_{\hat{\beta}_{4}}\left (V_{4}\right )}^{h'}_{h} \overline{R_{\hat{\gamma}_{1}}\left (W_{1}\right )}^{e'_{1}}_{e_{1}}  \overline{R_{\hat{\gamma}_{3}}\left (W_{3}\right )}^{g'_{1}}_{g_{1}} \sum_{\nu_{i}, \lambda_{i}} C_{\nu_{1}}\left (\hat{\beta}_{1}, \hat{\alpha}_{1}, \overline{\hat{\beta}}_{4}\right )^{e'a'}_{h} \,\times \\
        &\times C_{\nu_{2}}\left (\hat{\beta}_{4}, \hat{\alpha}_{4}, \overline{\hat{\beta}}_{3}\right )^{h'd'}_{g} \, \overline{C}_{\nu_{3}} \left (\hat{\beta}_{1}, \hat{\alpha}_{5}, \overline{\hat{\beta}}_{3}\right )^{g'}_{e j_{3}} \, C_{\lambda_{1}} \left (\hat{\gamma}_{1}, \hat{\alpha}_{5}, \overline{\hat{\gamma}}_{3}\right )^{e'_{1} j'_{3}}_{g_{1}} \, \times \\
        &\times \overline{C}_{\lambda_{2}} \left (\hat{\beta}_{2}, \hat{\alpha}_{3}, \overline{\hat{\gamma}_{3}}\right )^{g'_{1}}_{fc} \, \overline{C}_{\lambda_{3}}\left (\hat{\gamma}_{1}, \hat{\alpha}_{2}, \overline{\hat{\beta}}_{2}\right )^{f'}_{e_{1}b} \, .
    \end{split}
\end{equation}
We denoted the uncontracted indices by $\cdot$, just as we did in (\ref{eq: W4 exp value}), to compare the two expressions. This operator will be glued to the plaquettes as on the left of figure \ref{fig: gluing of a 3 and a 4-line junction}. \par
By fusion on the lattice we mean taking the holonomy to 1. When performing the integrals we then add the following projection

\begin{equation}
    \sum_{\hat{\alpha}_{5}, \hat{\gamma}_{1}, \hat{\gamma}_{3}} \dim\left (\hat{\gamma}_{1}\right ) \delta\left (U_{5}-1\right ) \delta_{\nu_{3}, \lambda_{1}} \, ,
    \label{eq: fusion relation}
\end{equation}
where $\nu_{3}, \lambda_{1}$ are the indices of the Clebsch-Gordan coefficients living at the ends of the holonomy that is going to 1. A composition followed by this projection corresponds to the fusion operation on $U_{5}$. We get

\begin{equation}
     \left <\mathbb{W}_{3} \circ_{U_{5}} \mathbb{W}_{3}\right> = \sum_{\hat{\alpha}_{5}} N^{\hat{\beta}_{3}}_{\hat{\beta}_{1} \hat{\alpha}_{5}} \left <\mathbb{W}_{4}\right> \, .
\end{equation}
This result can be reproduced using (\ref{eq: integral for fusion number}) for the $U$s; and (\ref{eq: schur orthogonality}) for the $V$s and $W$s. In particular, one of the extra $1/\dim$ factors from the integration of $W_{1}, W_{3}$ gets canceled by the one in (\ref{eq: fusion relation}) and the other one goes along with the Clebsch-Gordan coefficients containing $\hat{\alpha}_{5}$ into the fusion number (\ref{eq: fusion number from clebsch-gordan}). \par
This shows that by fusing two 3-line junctions we do not get a new 4-line junction, but a combination of 4-lines junctions that are already in the set. This is a local operation that only affects the holonomies adjacent to $U_{5}$, i.e. $V_{1},V_{2},W_{1},W_{3}$ and the Clebsch-Gordan coefficients involving $\hat{\alpha}_{5}$. Therefore, we can reasonably expect to be able to iterate the procedure to fuse an n- and an m-line junction into a linear combination of (n+m-2)-line junctions.

\subsection{Recovering the IRF crossing}
\label{section: IRF crossing}

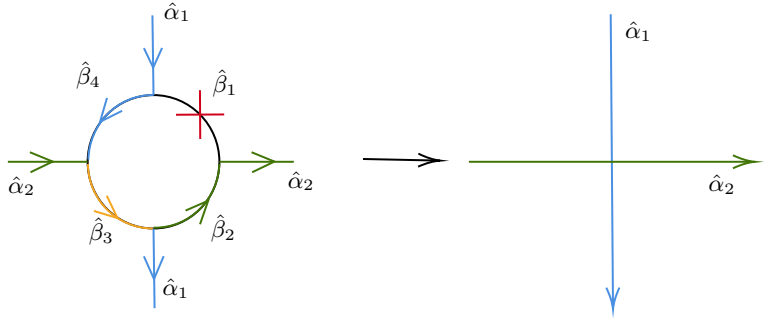
\begin{figure}
    \centering
    \tikzset{every picture/.style={line width=0.75pt}}

\begin{tikzpicture}[x=0.75pt,y=0.75pt,yscale=-1,xscale=1]

\draw   (81.09,132.83) .. controls (81.09,114.3) and (95.81,99.28) .. (113.97,99.28) .. controls (132.13,99.28) and (146.85,114.3) .. (146.85,132.83) .. controls (146.85,151.36) and (132.13,166.38) .. (113.97,166.38) .. controls (95.81,166.38) and (81.09,151.36) .. (81.09,132.83) -- cycle ;

\draw [color={rgb, 255:red, 74; green, 144; blue, 226 }  ,draw opacity=1 ]   (113.97,99.28) -- (113.43,59.18) ;

\draw [color={rgb, 255:red, 74; green, 144; blue, 226 }  ,draw opacity=1 ]   (114.5,206.47) -- (113.97,166.38) ;

\draw [color={rgb, 255:red, 65; green, 117; blue, 5 }  ,draw opacity=1 ]   (146.85,132.83) -- (184,132.83) ;

\draw [color={rgb, 255:red, 65; green, 117; blue, 5 }  ,draw opacity=1 ]   (81.09,132.83) -- (41.27,132.83) ;
\draw  [color={rgb, 255:red, 74; green, 144; blue, 226 }  ,draw opacity=1 ] (118.25,75.55) -- (113.81,85.44) -- (109.18,75.64) ;
\draw  [color={rgb, 255:red, 74; green, 144; blue, 226 }  ,draw opacity=1 ] (119.05,180.29) -- (114.22,191.81) -- (109.17,180.38) ;
\draw  [color={rgb, 255:red, 65; green, 117; blue, 5 }  ,draw opacity=1 ] (162.92,128.02) -- (174.31,132.72) -- (163.21,138.1) ;
\draw  [color={rgb, 255:red, 65; green, 117; blue, 5 }  ,draw opacity=1 ] (52.26,128.02) -- (63.65,132.72) -- (52.56,138.1) ;
\draw  [color={rgb, 255:red, 65; green, 117; blue, 5 }  ,draw opacity=1 ] (129.97,157.37) -- (141.25,152.42) -- (137.19,164.25) ;
\draw  [color={rgb, 255:red, 245; green, 166; blue, 35 }  ,draw opacity=1 ] (90.63,149.88) -- (95.99,161.15) -- (84.22,157.56) ;
\draw  [color={rgb, 255:red, 74; green, 144; blue, 226 }  ,draw opacity=1 ] (98.31,107.18) -- (87.29,112.73) -- (90.74,100.7) ;
\draw  [color={rgb, 255:red, 208; green, 2; blue, 27 }  ,draw opacity=1 ] (137.29,121.37) -- (137.15,96.82)(125.2,109.17) -- (149.25,109.02) ;

\draw  [draw opacity=0] (81.47,131.64) .. controls (81.86,113.7) and (96.07,99.28) .. (113.53,99.28) .. controls (113.68,99.28) and (113.83,99.28) .. (113.98,99.28) -- (113.53,132.42) -- cycle ; \draw  [color={rgb, 255:red, 74; green, 144; blue, 226 }  ,draw opacity=1 ] (81.47,131.64) .. controls (81.86,113.7) and (96.07,99.28) .. (113.53,99.28) .. controls (113.68,99.28) and (113.83,99.28) .. (113.98,99.28) ;  

\draw  [draw opacity=0] (146.83,133.85) .. controls (146.31,151.79) and (132,166.11) .. (114.54,165.98) .. controls (114.39,165.98) and (114.24,165.97) .. (114.09,165.97) -- (114.77,132.84) -- cycle ; \draw  [color={rgb, 255:red, 65; green, 117; blue, 5 }  ,draw opacity=1 ] (146.83,133.85) .. controls (146.31,151.79) and (132,166.11) .. (114.54,165.98) .. controls (114.39,165.98) and (114.24,165.97) .. (114.09,165.97) ;  

\draw  [draw opacity=0] (113.66,166.38) .. controls (96.08,166.37) and (81.65,152.22) .. (81.25,134.42) .. controls (81.24,134.27) and (81.24,134.12) .. (81.24,133.98) -- (113.71,133.65) -- cycle ; \draw  [color={rgb, 255:red, 245; green, 166; blue, 35 }  ,draw opacity=1 ] (113.66,166.38) .. controls (96.08,166.37) and (81.65,152.22) .. (81.25,134.42) .. controls (81.24,134.27) and (81.24,134.12) .. (81.24,133.98) ;  

\draw    (218.56,131.42) -- (255.05,132.2) ;
\draw [shift={(257.04,132.24)}, rotate = 181.22] [color={rgb, 255:red, 0; green, 0; blue, 0 }  ][line width=0.75]    (10.93,-3.29) .. controls (6.95,-1.4) and (3.31,-0.3) .. (0,0) .. controls (3.31,0.3) and (6.95,1.4) .. (10.93,3.29)   ;

\draw [color={rgb, 255:red, 74; green, 144; blue, 226 }  ,draw opacity=1 ]   (342.19,58.6) -- (343.32,205.06) ;
\draw [shift={(343.34,207.06)}, rotate = 269.56] [color={rgb, 255:red, 74; green, 144; blue, 226 }  ,draw opacity=1 ][line width=0.75]    (10.93,-3.29) .. controls (6.95,-1.4) and (3.31,-0.3) .. (0,0) .. controls (3.31,0.3) and (6.95,1.4) .. (10.93,3.29)   ;

\draw [color={rgb, 255:red, 65; green, 117; blue, 5 }  ,draw opacity=1 ]   (271.4,132.83) -- (412.13,132.83) ;
\draw [shift={(414.13,132.83)}, rotate = 180] [color={rgb, 255:red, 65; green, 117; blue, 5 }  ,draw opacity=1 ][line width=0.75]    (10.93,-3.29) .. controls (6.95,-1.4) and (3.31,-0.3) .. (0,0) .. controls (3.31,0.3) and (6.95,1.4) .. (10.93,3.29)   ;

\draw (117.7,50.95) node [anchor=north west][inner sep=0.75pt]  [font=\footnotesize]  {$\hat{\alpha }_{1}$};
 
\draw (116.9,190.05) node [anchor=north west][inner sep=0.75pt]  [font=\footnotesize]  {$\hat{\alpha }_{1}$};
 
\draw (179.44,136.05) node [anchor=north west][inner sep=0.75pt]  [font=\footnotesize]  {$\hat{\alpha }_{2}$};
 
\draw (39.92,138.5) node [anchor=north west][inner sep=0.75pt]  [font=\footnotesize]  {$\hat{\alpha }_{2}$};
 
\draw (141.95,84.95) node [anchor=north west][inner sep=0.75pt]  [font=\footnotesize]  {$\hat{\beta }_{1}$};
 
\draw (141.15,157.78) node [anchor=north west][inner sep=0.75pt]  [font=\footnotesize]  {$\hat{\beta }_{2}$};
 
\draw (80.21,159.41) node [anchor=north west][inner sep=0.75pt]  [font=\footnotesize]  {$\hat{\beta }_{3}$};
 
\draw (73.8,80.86) node [anchor=north west][inner sep=0.75pt]  [font=\footnotesize]  {$\hat{\beta }_{4}$};
 
\draw (347.83,60.77) node [anchor=north west][inner sep=0.75pt]  [font=\footnotesize]  {$\hat{\alpha }_{1}$};
 
\draw (389.53,137.68) node [anchor=north west][inner sep=0.75pt]  [font=\footnotesize]  {$\hat{\alpha }_{2}$};

\end{tikzpicture}

    \caption{Moves to get intersecting Wilson lines from a 4-line junction.}
    \label{fig: intersecting Wilson lines from 4-line junction}
\end{figure}

The IRF crossing was discussed by Witten in \cite{Witten:1991we} and it is realized as the intersection of two Wilson lines. To bring this into our notation, we define the following junction

\begin{equation}
    \mathbb{W}_{+} \left( \left \{ \hat{\alpha}_{1}, \hat{\alpha}_{2} \right\} \right)^{ab}_{a'b'} := R_{\hat{\alpha}_{1}}(U_{1})^{a}_{a'} R_{\hat{\alpha}_{2}}(U_{2})^{b}_{b'} \, .
\end{equation}

The local expectation value of the IRF crossing, obtained without specifying the topology far from it (right hand side of figure \ref{fig: intersecting Wilson lines from 4-line junction}) is

\begin{equation}
    \begin{split}
        \left < \mathbb{W}_{+} \right> =&
        \sum_{\hat{\rho}_{i}} \prod_{i} \dim\left (\hat{\rho}_{i}\right ) e^{-\frac{\epsilon_{i}}{2} c_{2}\left (\hat{\rho}_{i}\right )} \sum_{\mu_{i}}\epsilon_{\mu_{1}}\left (\hat{\rho}_{1}, \hat{\alpha}_{1}, \overline{\hat{\rho}}_{4}\right )^{a \cdot}_{h} \, \overline{\epsilon}_{\mu_{2}}\left (\hat{\rho}_{1}, \hat{\alpha}_{2}, \overline{\hat{\rho}}_{2}\right )^{c}_{b \cdot} \, \overline{\epsilon}_{\mu_{3}}\left (\hat{\rho}_{2}, \hat{\alpha}_{1}, \overline{\hat{\rho}}_{3}\right )^{e}_{d \cdot} \, \times \\
        &\times \epsilon_{\mu_{4}}\left (\hat{\rho}_{4}, \hat{\alpha}_{2}, \overline{\hat{\rho}}_{3}\right )^{g \cdot}_{f} R_{\hat{\rho}_{1}}\left (\mathcal{U}_{1}\right )^{b}_{a} R_{\hat{\rho}_{2}}\left (\mathcal{U}_{2}\right )^{d}_{c} R_{\hat{\rho}_{3}}\left (\mathcal{U}_{3}\right )^{f}_{e} R_{\hat{\rho}_{4}}\left (\mathcal{U}_{4}\right )^{h}_{g} R_{\hat{\alpha}_{1}}\left (\mathcal{U}_{5}\right )^{l}_{i} R_{\hat{\alpha}_{2}}\left (\mathcal{U}_{6}\right )^{n}_{m} \,  \times \\
        & \times \epsilon_{\mu_{2}}\left (\hat{\rho}_{1}, \hat{\alpha}_{2}, \overline{\hat{\rho}}_{2}\right )^{a' m'}_{c'} \,  \epsilon_{\mu_{3}}\left (\hat{\rho}_{2}, \hat{\alpha}_{1}, \overline{\hat{\rho}}_{3}\right )^{c' i'}_{e'} \, \overline{\epsilon}_{\mu_{1}}\left (\hat{\rho}_{1}, \hat{\alpha}_{1}, \overline{\hat{\rho}}_{4}\right )^{g'}_{a' i'} \, \overline{\epsilon}_{\mu_{4}}\left (\hat{\rho}_{4}, \hat{\alpha}_{2}, \overline{\hat{\rho}}_{3}\right )^{e'}_{g' m'} \, .
    \end{split}
    \label{eq: final result intersecting lines}
\end{equation}
Following the same conventions of the previous section, we denoted $\cdot$ the uncontracted indices of the junction. Where the last four 3j symbols rearrange into a 6j symbol. Let us restrict for simplicity to simply reducible groups (see Appendix \ref{appendix: 6j}), the result can be written as

\begin{equation}
\begin{split}  
    \left < \mathbb{W}_{+} \right> =&
        \sum_{\hat{\rho}_{i}} \prod_{i} \dim\left (\hat{\rho}_{i}\right ) \epsilon \left (\hat{\rho}_{1}, \hat{\alpha}_{1}, \hat{\rho}_{4}\right )^{ai}_{h} \, \overline{\epsilon}\left (\hat{\rho}_{1}, \hat{\alpha}_{2}, \hat{\rho}_{2}\right )^{c}_{bn} \, \overline{\epsilon}\left (\hat{\rho}_{2}, \hat{\alpha}_{1}, \hat{\rho}_{3}\right )^{e}_{dl} \, \epsilon\left (\hat{\rho}_{4}, \hat{\alpha}_{2}, \hat{\rho}_{3}\right )^{gm}_{f} \, \times \\
        &\times R_{\hat{\rho}_{1}}\left (\mathcal{U}_{1}\right )^{b}_{a} R_{\hat{\rho}_{2}}\left (\mathcal{U}_{2}\right )^{d}_{c} R_{\hat{\rho}_{3}}\left (\mathcal{U}_{3}\right )^{f}_{e} R_{\hat{\rho}_{4}}\left (\mathcal{U}_{4}\right )^{h}_{g} R_{\hat{\alpha}_{1}}\left (\mathcal{U}_{5}\right )^{l}_{i} R_{\hat{\alpha}_{2}}\left (\mathcal{U}_{6}\right )^{n}_{m} e^{-\frac{\epsilon_{i}}{2} c_{2}\left (\hat{\rho}_{i}\right )} \times \\
        &\times \left (-1\right )^{\hat{\rho}_{4} - \hat{\rho}_{2} + \hat{\alpha}_{1} - \hat{\alpha}_{2}} \begin{pmatrix}
            \hat{\alpha}_{2} & \hat{\rho}_{3} & \hat{\rho}_{4} \\
            \hat{\alpha}_{1} & \hat{\rho}_{1} & \hat{\rho}_{2}
        \end{pmatrix}  \, .
\end{split}
\label{eq: IRF SR groups}
\end{equation}
Now we compute the expectation value of a 4-line junction inserting

\begin{equation}  
    p = \sum_{\hat{\beta}_{i}, \hat{\alpha}_{3}, \hat{\alpha}_{4}} \left (-1\right )^{\hat{\beta}_{3} + 2 \hat{\rho}_{2} -\hat{\alpha}_{1} + \hat{\alpha}_{2}} \dim\left (\hat{\beta}_{5}\right )\chi_{\overline{\hat{\beta}}_{5}} \left (V_{1}V_{2}V_{3}V_{4}\right ) \delta_{\hat{\beta}_{1}, \hat{e}} \delta_{\hat{\alpha}_{1}, \hat{\alpha}_{3}} \, \delta_{\hat{\alpha}_{2}, \hat{\alpha}_{4}} \, ,
\end{equation}
corresponding to the moves shown in the left of figure \ref{fig: intersecting Wilson lines from 4-line junction}. We get

\begin{equation}
    \left < p \mathbb{W}_{4} \right > =  \left < \mathbb{W}_{+} \right> \, .
\end{equation}

Since this is the most involved computation in this work, let us show some intermediate steps. If we forget for a second the phase factor in $p$, that is specific for simply reducible groups, we can sketch the computation for general compact groups. We integrate out the $U$s using (\ref{eq: integral for fusion number}). For the $V$s we first rewrite the $\overline{\hat{\beta}}_{5}$ terms in $p$ (\ref{eq: unitarity}) and then use (\ref{eq: fusion number variation}). Setting $\hat{\alpha}_{1} = \hat{\alpha}_{3}, \hat{\alpha}_{2} = \hat{\alpha}_{4}$ and $\hat{\beta}_{1}$ to be trivial we end up with

\begin{equation} 
    \begin{split}
        \left <p \mathbb{W}_{4}\right> =& \sum_{\hat{\rho}_{i}} \prod_{i} \dim\left (\hat{\rho}_{i}\right ) e^{-\frac{\epsilon_{i}}{2} c_{2}\left (\hat{\rho}_{i}\right )} \sum_{\mu_{i}} \epsilon_{\mu_{1}}\left (\hat{\rho}_{1}, \hat{\alpha}_{1}, \overline{\hat{\rho}}_{4}\right )^{i x}_{p'}  \overline{\epsilon}_{\mu_{1}}\left (\hat{\rho}_{1}, \hat{\alpha}_{1}, \overline{\hat{\rho}}_{4}\right )^{p}_{i' a'}  \epsilon_{\mu_{2}}\left (\hat{\rho}_{1}, \hat{\alpha}_{2}, \overline{\hat{\rho}}_{2}\right )^{j b}_{k'}  \times \\
    &\times \overline{\epsilon}_{\mu_{2}}\left (\hat{\rho}_{1}, \hat{\alpha}_{2}, \overline{\hat{\rho}}_{2}\right )^{k}_{j' x}  \epsilon_{\mu_{3}}\left (\hat{\rho}_{2}, \hat{\alpha}_{1}, \overline{\hat{\rho}}_{3}\right )^{l c}_{m'}  \overline{\epsilon}_{\mu_{3}}\left (\hat{\rho}_{2}, \hat{\alpha}_{1}, \overline{\hat{\rho}}_{3}\right )^{m}_{l' x}  \epsilon_{\mu_{4}}\left (\hat{\rho}_{4}, \hat{\alpha}_{2}, \overline{\hat{\rho}}_{3}\right )^{o x}_{n'}  \times \\
    &\times \overline{\epsilon}_{\mu_{4}}\left (\hat{\rho}_{4}, \hat{\alpha}_{2}, \overline{\hat{\rho}}_{3}\right )^{n}_{o' d'} R_{\hat{\rho}_{1}}\left (\mathcal{U}_{1}\right )^{j'}_{i} R_{\hat{\rho}_{2}}\left (\mathcal{U}_{2}\right )^{l'}_{k} R_{\hat{\rho}_{3}}\left (\mathcal{U}_{3}\right )^{n'}_{m} R_{\hat{\rho}_{4}}\left (\mathcal{U}_{4}\right )^{p'}_{o} \sum_{\hat{\beta}_{2}, \hat{\beta}_{3}, \hat{\beta}_{4}, \hat{\beta}_{5}} \dim\left (\hat{\beta}_{5}\right )  \times \\
    & \times \overline{\epsilon}_{\mu_{5}}\left (\overline{\hat{\rho}}_{1}, \hat{\beta}_{5}, \hat{e}\right )^{i'}_{e'q}  \epsilon_{\mu_{5}}\left (\overline{\hat{\rho}}_{1}, \hat{\beta}_{5}, \hat{e}\right )^{e q'}_{j}  \overline{\epsilon}_{\mu_{6}}\left (\overline{\hat{\rho}}_{2}, \hat{\beta}_{5}, \hat{\beta}_{2}\right )^{k'}_{f'q'}  \epsilon_{\mu_{6}}\left (\overline{\hat{\rho}}_{2}, \hat{\beta}_{5}, \hat{\beta}_{2}\right )^{fr}_{l} \times \\
    &\times \overline{\epsilon}_{\mu_{7}}\left (\overline{\hat{\rho}}_{3}, \hat{\beta}_{5}, \hat{\beta}_{3}\right )^{m'}_{g'r}  \epsilon_{\mu_{7}}\left (\overline{\hat{\rho}}_{3}, \hat{\beta}_{5}, \hat{\beta}_{3}\right )^{gr'}_{n}  \overline{\epsilon}_{\mu_{8}}\left (\overline{\hat{\rho}}_{4}, \hat{\beta}_{5}, \hat{\beta}_{4}\right )^{o'}_{h'r'}  \epsilon_{\mu_{8}}\left (\overline{\hat{\rho}}_{4}, \hat{\beta}_{5}, \hat{\beta}_{4}\right )^{h q}_{p}  \times\\  
    & \times C_{\nu_{1}}\left (\hat{e}, \hat{\alpha}_{1}, \overline{\hat{\beta}}_{4}\right )^{e'a'}_{h}  \overline{C}_{\nu_{2}}\left (\hat{e}, \hat{\alpha}_{2}, \overline{\hat{\beta}}_{2}\right )^{f'}_{eb}  \overline{C}_{\nu_{3}} \left (\hat{\beta}_{2}, \hat{\alpha}_{1}, \overline{\hat{\beta}}_{3}\right )^{g'}_{f c}  C_{\nu_{4}} \left (\hat{\beta}_{4}, \hat{\alpha}_{2}, \overline{\hat{\beta}}_{3}\right )^{h' d'}_{g} \, .
    \end{split}
\end{equation}
When solving $\epsilon_{\mu_{5}}, \overline{\epsilon}_{\mu_{5}}$ using (\ref{eq: special case}) we get an extra factor $1/\dim\left (\hat{\beta}_{5}\right )$ that cancels with the $\dim\left (\hat{\beta}_{5}\right )$ in $p$. $C_{\nu_{1}}, \overline{C}_{\nu_{2}}$ can be solved using (\ref{eq: special case for CG}). Doing that, we end up with the following

\begin{equation}
    \epsilon_{\mu_{2}}\left (\hat{\rho}_{1}, \hat{\alpha}_{2}, \overline{\hat{\rho}}_{2}\right )^{jb}_{k'} \, \overline{\epsilon}_{\mu_{6}}\left (\hat{\rho}_{1}, \hat{\alpha}_{2}, \overline{\hat{\rho}}_{2}\right )_{jb}^{k'} = \delta_{\mu_{2}, \mu_{6}} \, ; \hspace{5mm}  \epsilon_{\mu_{8}}\left (\hat{\rho}_{1}, \hat{\alpha}_{1}, \overline{\hat{\rho}}_{4}\right )^{hq}_{p} \, \overline{\epsilon}_{\mu_{1}}\left (\hat{\rho}_{1}, \hat{\alpha}_{1}, \overline{\hat{\rho}}_{4}\right )_{hq}^{p} = \delta_{\mu_{8}, \mu_{1}} \, ,
    \label{eq: intermediate step IRF}
\end{equation}
where we used (\ref{eq: CG contraction}). 
We then have to turn $\overline{C}_{\nu_{3}}, C_{\nu_{4}}$ into 3j-symbols by adding the appropriate normalization so that $\epsilon_{\mu_{3}}, \epsilon_{\mu_{2}}, \overline{\epsilon}_{\nu_{3}}, \overline{\epsilon}_{\mu_{7}}$ and $\epsilon_{\mu_{7}}, \epsilon_{\nu_{4}}, \overline{\epsilon}_{\mu_{4}}, \overline{\epsilon}_{\mu_{1}}$ can assemble into two 6j symbols. \par
Let us restrict to simply reducible groups to compare to the result (\ref{eq: IRF SR groups}). With this simplification, (\ref{eq: intermediate step IRF}) no longer has $\mu$ indices and gives the fusion number which we take to be 1, because if it was 0 the whole expectation value would be trivial anyways. Reintroducing the phase factor contained in $p$ and using the properties of simply reducible groups (see Appendix \ref{appendix: 6j}), 8 3j symbols assemble into 2 6j symbols and we end up with

\begin{equation}
    \begin{split}
        \left <p \mathbb{W}_{4}\right> =& \sum_{\hat{\rho}_{i}} \prod_{i} \dim\left (\hat{\rho}_{i}\right ) \epsilon\left (\hat{\rho}_{1}, \hat{\alpha}_{1}, \hat{\rho}_{4}\right )^{i x}_{p'} \, \overline{\epsilon}\left (\hat{\rho}_{1}, \hat{\alpha}_{2}, \hat{\rho}_{2}\right )^{k}_{j' x} \, \overline{\epsilon} \left (\hat{\rho}_{2}, \hat{\alpha}_{1}, \hat{\rho}_{3}\right )^{m}_{l' x} \, \epsilon\left (\hat{\rho}_{4}, \hat{\alpha}_{2}, \hat{\rho}_{3}\right )^{o x}_{n'} \, \times \\
        & \times R_{\hat{\rho}_{1}}\left (\mathcal{U}_{1}\right )^{j'}_{i} R_{\hat{\rho}_{2}}\left (\mathcal{U}_{2}\right )^{l'}_{k} R_{\hat{\rho}_{3}}\left (\mathcal{U}_{3}\right )^{n'}_{m} R_{\hat{\rho}_{4}}\left (\mathcal{U}_{4}\right )^{p'}_{o} \, e^{-\frac{\epsilon_{i}}{2} c_{2}\left (\hat{\rho}_{i}\right )} \times \\
        & \times \left (-1\right )^{\hat{\rho}_{4} -\hat{\rho}_{2} + \hat{\alpha}_{1} - \hat{\alpha}_{2}} \sum_{\hat{\beta}_{3}} \dim \hat{\beta}_{3} \left (-1\right )^{\hat{\beta}_{3} + \hat{\rho}_{4} + \hat{\rho}_{2}} \begin{pmatrix}
        \hat{\rho}_{3} & \hat{\alpha}_{2} & \hat{\rho}_{4} \\
        \hat{\alpha}_{1} & \hat{\rho}_{1} & \hat{\beta}_{3}
    \end{pmatrix}
    \begin{pmatrix}
        \hat{\alpha}_{2} & \hat{\alpha}_{1} & \hat{\beta}_{3} \\
        \hat{\rho}_{3} & \hat{\rho}_{1} & \hat{\rho}_{2}
    \end{pmatrix} \, .
    \end{split}
\end{equation}
Now we can use (\ref{eq: 6j symbol identity}) and get precisely (\ref{eq: IRF SR groups}).

\section{Discussion}

We have introduced a specific refinement of the lattice model for 2dYM. We showed that the refinement reproduced results regarding the partition function and expectation values of defects and defect networks from the vast literature on 2dYM, while maintaining subdivision invariance: the essential property allowing for exact computations. We further showed that a variety of known defects of the continuum theory, including networks of defects of different dimensions, can be realized as functions of the classical degrees of freedom in this refinement with all locality properties intact. This was our main criterion for a successful refinement. Finally, we demonstrated that our collection of defects was complete with respect to fusion.

In section \ref{section: IRF crossing}, we recovered the IRF crossing, concentrating, for simplicity, on simply reducible groups. The generalizations to arbitrary compact groups, presumably requires the use of identities for $3nj$-symbols appearing in e.g. \cite{Derome,Butler1975}. We believe that the result still holds, possibly with different phase factor in the weight (see Appendix \ref{appendix: 6j}).
Whereas in section \ref{section: degeneration} we deal with operator equations, we were only able to perform the fusion (section \ref{section: fusion}) and the IRF check (section \ref{section: IRF crossing}) in terms of expectation values. This is unusual and perhaps a sign that there is a more refined prescription for the defects than the one outlined here.

The approach taken in \cite{Iraso:2018mwa} produced similar results in the continuum, but did not address the issue of defect networks. We would further argue in favor of our approach by noting that it works for arbitrary genus, structure group, and theta angle, and also for additional non-topological deformations of the theory, such as including higher Casimirs in the action \cite{Witten:1992xu,Ganor:1994bq,RUSAKOV1994258,Griguolo:2021rke}, as long as the Hamiltonian stays diagonal in the representation basis. However, our refinement was introduced ad hoc, while the additional degrees of freedom in \cite{Iraso:2018mwa} were the result of applying the BV formalism. We believe that the two approaches are complementary, with the BV formalism providing, at least on the Lie algebra level, an appropriate guide for the required degrees of freedom living on vertices. It should also be noted that at least some of these degrees of freedom are already visible in the cohomological supersymmetric version of topological 2dYM discussed in \cite{Witten:1992xu}.

2dYM admits 0, 1 and 2 dimensional defects: Wilson points, Wilson lines and discrete theta angles. The points are vortex or monodromy operators, a 2d analog of Gukov-Witten surface operators in 4d\footnote{Incidentally, we mention that recently \cite{Choi:2024ktc} a class of Gukov-Witten surface operators in four dimensional $\mathcal{N}=4$ has been shown to be effectively described by an operator insertion in two-dimensional Yang-Mills, exploiting the localization properties of the $\mathbb{S}^2$ 1/8 BPS subsector. It would be nice to understand if this insertion could be constructed from our formalism. More generally it would be interesting to characterize the $\mathcal{N}=4$ observables computed by 2dYM from the point of view developed here.} \cite{Gukov:2006jk,Gukov:2008sn}. The analog of these points in 3d has been studied in \cite{Witten:1988hf,Moore:1989yh} and their supersymmetric version has been studied in 2d \cite{Okuda:2015yra,Hosomichi:2017dbc,Hosomichi:2015pia} and in 3d \cite{Drukker:2012sr,Kapustin:2012iw,Hosomichi:2021gxe,Drukker:2023bip}. Some 3d gauge theories also admit mixed Wilson-vortex line operators \cite{Griguolo:2021rke}. It would be interesting to recover these operators enriching the lattice description of 3d gauge theory. A first step in this direction would be to study defects using the  lattice formulation of pure Chern-Simons theory, for which a number of proposals exist, see the recent paper \cite{Jacobson:2024hov} and reference therein. 

Two-dimensional Yang-Mills theory admits also a large-N description in terms of the so-called Gross-Taylor string theory \cite{Gross:1993hu,Gross:1993yt}, counting maps from genus g worldsheets into the Riemann surface where the gauge theory lives. Recently, it was argued that a one-form symmetry acts on the relevant string sigma model \cite{Cordes:1994fc,Horava:1995ic}, providing a decomposition for the Gross-Taylor theory \cite{Sharpe:2022ene,Pantev:2023dim}. It would be interesting to better understand the emergence of such symmetry at large N and to study line defects in the string set-up.

Our refinement enables us to control the 1-form symmetry gauging procedure plaquette by plaquette, and could perhaps be applied to generalized gauging procedures \cite{Roumpedakis:2022aik}. In (\ref{eq: half gauging}) we performed a half gauging. What we get can be interpreted as an interface defect between the gauged and ungauged theory, a type of condensation defect \cite{Gaiotto:2019xmp}. Such interfaces were discussed, for instance, in \cite{Carqueville:2023jhb} in the context of the Ising model. \par

\appendix

\section{Harmonic Analysis on Compact Groups}
\label{appendix: harmonic analysis}

The following results can be found in standard textbook treatments, such as the first chapter of \cite{Vilenkin1968SpecialFA} or \cite{Sepanski2006CompactLG}. We list here for convenience the formulae that we use throughout. 

Let $G$ be a \textit{compact} Lie group. We have a few immediate consequences of its compactness: all representations are finite dimensional; every representation is unitary with respect to an inner product; we can define a left and right invariant measure called Haar measure $dg$; we can do harmonic analysis on $G$. Specifically, the Haar measure has the following properties:
\begin{enumerate}
    \item Left and right invariance: $d\left (gh\right ) = d\left (hg\right ) = dg$, and, as a consequence,

    \begin{equation}
        \int dg f\left (gh\right ) = \int dg f\left (g\right ) \, ;
        \label{eq: Haar property 1}
    \end{equation}

    \item Normalization condition:

    \begin{equation}
        \int dg = 1 \, .
        \label{eq: normalisation haar measure}
    \end{equation}
\end{enumerate}

Let Greek indices label the set of inequivalent irreducible unitary representations of $G$. Let also $R_{\hat{\alpha}}\left (g\right )^{i}_{j}$ denote the matrix elements of the $\hat{\alpha}$ representation of the group element $g$. We consider only unitary representations, so that
\begin{equation}
    \overline{R_{\hat{\alpha}}\left (U\right )}^{i}_{j} = R_{\overline{\hat{\alpha}}} \left (U\right )^{i}_{j} = R_{\hat{\alpha}}\left (U^{-1}\right )^{j}_{i} \, ,
    \label{eq: unitarity}
\end{equation}
and
\begin{equation}
\overline{\overline{R_{\hat{\alpha}}\left (U\right )}}^{i}_{j} = R_{\hat{\alpha}}\left (U\right )^{i}_{j}
\end{equation}
using Vilenkin's conventions \cite{Vilenkin1968SpecialFA}.

By the Peter-Weyl theorem, the set $\dim\left (\hat{\alpha}\right ) R_{\hat{\alpha}}\left (g\right )^{i}_{j}$ forms a complete orthonormal system for the invariant measure $dg$ on $G$, and any square-integrable function $f\left (g\right ) \in L^{2}\left (G\right )$ can be decomposed in this basis. This implies

\begin{equation}
    \int_{G} dg R_{\hat{\alpha}}\left (g\right )^{i}_{j} \overline{R_{\hat{\beta}}\left (g\right )^{k}_{l}} = \frac{\delta_{\hat{\alpha}, \hat{\beta}}}{\dim\left (\hat{\alpha}\right )} \cdot \delta_{i,k} \cdot \delta_{j,l} \, .
    \label{eq: schur orthogonality}
\end{equation}
$f\left (g\right )$ is called a \textit{class} function if 
\begin{equation}
   f\left (h^{-1}gh\right ) = f\left (g\right )\, , \quad  \forall h,g \in G\, .
\end{equation}
Every square-integrable class function can be decomposed in a Fourier series in the basis of characters. We have the following representation of the delta function using the characters \cite{Witten:1991we}

\begin{equation}
    \sum_{\hat{\alpha}} \chi_{\hat{\alpha}}\left (1\right ) \chi_{\hat{\alpha}} \left (U\right ) = \sum_{\hat{\alpha}} \dim\left (\hat{\alpha}\right ) \chi_{\hat{\alpha}}\left (U\right ) = \delta\left (U-1\right ) \, .
    \label{eq: basis rel}
\end{equation}
We can go between the point-wise and character representation of a class function as follows

\begin{equation}
    \begin{split}
        F: & \, \text{hol} \longrightarrow \text{char} \\
        & f\left (U\right ) \mapsto \int dU f\left (U\right ) \chi_{\hat{\alpha}}\left (U\right )
    \end{split}
\end{equation}

\begin{equation}
    \begin{split}
        \hat{F}: & \,  \text{char} \longrightarrow \text{hol}\\
        & f\left (\hat{\alpha}\right ) \mapsto \sum_{\hat{\alpha}} f\left (\hat{\alpha}\right ) \chi_{\hat{\alpha}}\left (U\right )
    \end{split}
\end{equation}
which is a Fourier type transform on $G$.

\section{Clebsch-Gordan coefficients and 3j symbols}
\label{appendix: clebsch-gordan}

We list here for simplicity some results we need for the computations. We will be focusing on general compact groups, sometimes specializing to simply reducible groups (see Appendix \ref{appendix: 6j}). References for simply reducible groups are \cite{3j6j,Wigner1993}, chapter 3 section 8 in \cite{Vilenkin1968SpecialFA} and Appendix C in \cite{Messiah1961QuantumMV}. For general compact groups, all details can be found in \cite{Derome,Butler1975}. \par

The tensor product of irreducible representations of $G$ can be decomposed as a sum of irreducible representations

\begin{equation}\hat{\alpha} \otimes \hat{\beta} = \bigoplus_{\hat{\gamma}} N^{\hat{\gamma}}_{\hat{\alpha} \hat{\beta}} \hat{\gamma}\, .
\end{equation}
$N$ is called fusion number and counts the number of times the representation $\hat{\gamma}$ appears in the decomposition of $\hat{\alpha} \otimes \hat{\beta}$. Fusion numbers for simply reducible groups are always either 0 or 1, but can be any integer for general compact groups. For the fusion number to be non-zero, we have the following necessary but not sufficient condition for the $N$-alities of $SU\left (N\right )$ representations
\begin{equation}
    N\text{-ality}\left (\hat{\alpha}\right ) + N\text{-ality}\left (\hat{\beta}\right ) + N\text{-ality}\left (\hat{\gamma}\right ) = 0 \, \, \left (\text{mod} \, N\right ) \, ,
\end{equation}
and a similar condition for any compact group.

The fusion numbers $N^{\hat{\gamma}}_{\hat{\alpha} \hat{\beta}}$, also denoted $H^{\hat{\gamma}}_{\hat{\alpha} \hat{\beta}}$, are the dimensions of the spaces of homomorphisms
\begin{equation}
H^{\hat{\gamma}}_{\hat{\alpha} \hat{\beta}} = \dim\left (\Hom\left (\hat{\alpha} \otimes \hat{\beta}, \hat{\gamma}\right )\right ) \simeq \dim\left (\Hom\left (\hat{\alpha} \otimes \hat{\beta}\otimes \overline{\hat{\gamma}} ,\hat e \right )\right ) \, .
\end{equation}
We can write this decomposition of the tensor product in terms of basis vectors for the representation vector spaces using using Clebsch-Gordan coefficients
\begin{equation}
    e_{\hat{\alpha}}^{i} \otimes e_{\hat{\beta}}^{j} = \sum_{\hat{\gamma}, k} C_{\mu_{\gamma}} \left (\hat{\alpha}, \hat{\beta}, \overline{\hat{\gamma}}\right )^{ij}_{k} \, e_{\hat{\gamma}}^{k} \, .
    \label{eq: clebsch}
\end{equation}
$C \left (\hat{\alpha}, \hat{\beta}, \overline{\hat{\gamma}}\right )^{ij}_{k}$ can be thought of as $G$-invariant basis tensors for the vector space $\hat{\alpha} \otimes \hat{\beta} \otimes \overline{\hat{\gamma}}$ and $C \left (\hat{\alpha}, \hat{\beta}, \hat{\gamma}\right )^{ijk}$ would be the $G$-invariant tensor in $\hat{\alpha} \otimes \hat{\beta} \otimes \hat{\gamma}$. Clebsch-Gordan coefficients are change of basis unitary matrices normalized as 
\[\sum_{\mu_{\gamma}} C_{\mu_{\gamma}} \left (\hat{\alpha}, \hat{\beta}, \overline{\hat{\gamma}}\right )^{k}_{ij} \overline{C}_{\mu_{\gamma}} \left (\hat{\alpha}, \hat{\beta}, \overline{\hat{\gamma}}\right )_{k}^{ij} = 1\,.\]

The index $\mu = 1,..., \dim\left (\hat{\alpha} \otimes \hat{\beta} \otimes \overline{\hat{\gamma}}\right )$ labels all Clebsch-Gordan coefficients appearing in the full decomposition of $\hat{\alpha} \otimes \hat{\beta}$; the subindex $\gamma$ indicates to which representation in the decomposition the coefficient is related. If we now fix a representation $\hat{\gamma}$, we can omit the subindex. With the chosen normalization, the coefficients are related to fusion numbers as

\begin{equation}
    \sum_{\mu} \frac{1}{\dim\left (\hat{\gamma}\right )} C_{\mu} \left (\hat{\alpha}, \hat{\beta}, \overline{\hat{\gamma}}\right )^{ij}_{k} \, \overline{C}_{\mu} \left (\hat{\alpha}, \hat{\beta}, \overline{\hat{\gamma}}\right )^{k}_{ij} = N^{\hat{\gamma}}_{\hat{\alpha} \hat{\beta}} \, . 
    \label{eq: fusion number from clebsch-gordan}
\end{equation}
These are the coefficients as they appear in \cite{Iraso:2018mwa}. Following \cite{Witten:1991we}, we will instead use the Wigner 3j symbols 

\begin{equation}
    \epsilon_{\mu} \left (\hat{\alpha}, \hat{\beta}, \overline{\hat{\gamma}}\right )^{ij}_{k} = \frac{C_{\mu}\left (\hat{\alpha}, \hat{\beta}, \overline{\hat{\gamma}}\right )^{ij}_{k}}{\left (\dim\left (\hat{\gamma}\right )\right )^{1/2}}\, ,
\end{equation}
These 3j symbols appear also in \cite{Butler1975}, where they are denoted as $\left (\alpha, \beta, \gamma^{\ast}\right )^{\mu i j}_{\quad k}$. \par

In terms of the representation matrices and Haar measure, we have

\begin{equation}
    \int dg R_{\hat{\alpha}}\left (g\right )^{i}_{i'} R_{\hat{\beta}} \left (g\right )^{j}_{j'} R_{\hat{\gamma}} \left (g^{-1}\right )^{k}_{k'} = \sum_{\mu} \epsilon_{\mu} \left (\hat{\alpha}, \hat{\beta}, \overline{\hat{\gamma}}\right )^{ij}_{k'}  \, \overline{\epsilon}_{\mu} \left (\hat{\alpha}, \hat{\beta}, \overline{\hat{\gamma}}\right )^{k}_{i'j'} \, .
    \label{eq: integral for fusion number}
\end{equation}
Note that our normalization for the Haar measure is different from e.g. eq. (5.12) in \cite{Butler1975} (cf eq. (\ref{eq: normalisation haar measure})). A variation of (\ref{eq: integral for fusion number}) which we will need is given by

\begin{equation}
\begin{split}
    \int dg R_{\hat{\alpha}}\left (g\right )^{i}_{i'} R_{\hat{\beta}} \left (g^{-1}\right )^{j}_{j'} R_{\hat{\gamma}} \left (g^{-1}\right )^{k}_{k'} =& \sum_{\mu} \epsilon_{\mu} \left (\hat{\alpha}, \overline{\hat{\beta}}, \overline{\hat{\gamma}}\right )^{i}_{j'k'}  \, \overline{\epsilon}_{\mu} \left (\hat{\alpha}, \overline{\hat{\beta}}, \overline{\hat{\gamma}}\right )^{jk}_{i'} \\
    =& \sum_{\mu} \overline{\epsilon}_{\mu} \left (\overline{\hat{\alpha}}, \hat{\beta}, \hat{\gamma}\right )^{i}_{j'k'} \, \epsilon_{\mu}\left (\overline{\hat{\alpha}}, \hat{\beta}, \hat{\gamma}\right )^{jk}_{i'} \, ,
    \label{eq: fusion number variation}
\end{split}
\end{equation}
where in the last equality we are using the following property (cf eq. 8.2 in \cite{Butler1975} and discussion below)

\begin{equation}
     \overline{\epsilon}_{\mu} \left (\hat{\alpha}, \hat{\beta}, \overline{\hat{\gamma}}\right )^{k}_{ij} = \epsilon_{\mu} \left (\overline{\hat{\alpha}}, \overline{\hat{\beta}}, \hat{\gamma}\right )^{k}_{ij} \, .
    \label{eq: conjugated clebsch-gordan}
\end{equation}

In the special case where one of the representations involved is the trivial, 3j symbols satisfy

\begin{equation}
    \epsilon_{\mu} \left (\hat{\alpha}, \hat{e}, \overline{\hat{\gamma}}\right )^{ij}_{k} = \frac{\delta_{\hat{\alpha},\hat{\gamma}} \cdot \delta_{i,k}}{\left (\dim\left (\hat{\gamma}\right )\right )^{1/2}} \hspace{10mm} \epsilon_{\mu} \left (\hat{\alpha}, \hat{\beta}, \hat{e}\right )^{ij}_{k} = \frac{\delta_{\hat{\alpha},\overline{\hat{\beta}}} \cdot \delta_{i,j}}{\left (\dim\left (\hat{\alpha}\right )\right )^{1/2}} \, .
    \label{eq: special case}
\end{equation}
These relations can be derived as follows

\begin{equation}
\begin{split}
    \int dg R_{\hat{\alpha}}\left (g\right )^{i}_{i'}R_{\hat{\beta}}\left (g\right )^{j}_{j'} =& \int dg R_{\hat{\alpha}}\left (g\right )^{i}_{i'}\overline{R_{\overline{\hat{\beta}}}\left (g\right )}^{j}_{j'} \\
    =& \frac{\delta_{\hat{\alpha}, \overline{\hat{\beta}}}}{\dim\left (\hat{\alpha}\right )} \cdot \delta_{i,j} \cdot \delta_{i',j'}\\
    =& \sum_{\mu} \epsilon_{\mu}\left (\hat{\alpha}, \hat{\beta}, \hat{e} \right )^{ij}_{k'} \, \overline{\epsilon}_{\mu}\left (\hat{\alpha}, \hat{\beta}, \hat{e} \right )^{k}_{i'j'} \, ,
\end{split}
\end{equation}
where we used (\ref{eq: schur orthogonality}), (\ref{eq: unitarity}) and (\ref{eq: integral for fusion number}). This also implies the following for Clebsch-Gordan coefficients
\begin{equation}
    C_{\mu}\left (\hat{\alpha}, \hat{e}, \overline{\hat{\gamma}}\right )^{ij}_{k'} = \delta_{\hat{\alpha}, \hat{\gamma}} \cdot \delta_{i, k'} \hspace{10mm} C_{\mu}\left (\hat{\alpha}, \hat{\beta}, \hat{e} \right )^{ij}_{k'} = \delta_{\hat{\alpha}, \overline{\hat{\gamma}}} \cdot \delta_{i,j} \, .
    \label{eq: special case for CG}
\end{equation}

\subsection{6j symbols and the IRF crossing}
\label{appendix: 6j}

In the IRF crossing, 6j symbols also appear.
For the sake of simplicity, we restrict this computation to simply reducible groups and use Wigner's notation \cite{Wigner1993}. The same computation can be done for generic compact groups by keeping track of additional phase factors. Simply reducible groups have fusion numbers $0$ or $1$, and every representation is equivalent to its complex conjugate. Every representation may be indexed by an integer or half integer label, and the example to have in mind is that of $SU\left (2\right )$. Hence, we can suppress the $\mu$ index and no longer keep track of complex conjugate representations, but we do keep track of covariant and contravariant indices. The integral (\ref{eq: integral for fusion number}) in Wigner's notation would give

\begin{equation}
    \epsilon\left (\hat{\alpha}, \hat{\beta}, \hat{\gamma}\right )^{ij}_{\quad k'} \, \overline{\epsilon}\left (\hat{\alpha}, \hat{\beta}, \hat{\gamma}\right )_{i' j'}^{\quad k} \, .
\end{equation}

We will use the following properties of the 3j symbols:

\begin{enumerate}
    \item If $\hat{\gamma}$ appears in $\hat{\alpha} \otimes \hat{\beta}$, we have

    \begin{equation}
        \left (-1\right )^{2 \hat{\alpha} + 2 \hat{\beta} + 2 \hat{\gamma}} = 1 \, .
    \end{equation}
    
    \item 3j symbols are symmetric under even permutations

    \begin{equation}
        \epsilon\left (\hat{\alpha}, \hat{\beta}, \hat{\gamma}\right )^{i j k} = \epsilon\left (\hat{\beta}, \hat{\gamma}, \hat{\alpha}\right )^{j k i} = \epsilon\left (\hat{\gamma}, \hat{\alpha}, \hat{\beta}\right )^{k i j} \, .
        \label{eq: even permutation}
    \end{equation}

    \item 3j symbols pick up a phase factor under odd permutations \footnote{For a general compact group we get phase factor for both even and odd permutations (cf eq. (6.2) in \cite{Butler1975}).}

    \begin{equation}
        \epsilon\left (\hat{\alpha}, \hat{\beta}, \hat{\gamma}\right )^{i j k} = \left (-1\right )^{\hat{\alpha} + \hat{\beta} + \hat{\gamma}} \epsilon\left (\hat{\beta}, \hat{\alpha}, \hat{\gamma}\right )^{j i k} \, .
    \end{equation}

    \item A 3j symbol with all contravariant indices is equivalent to the complex conjugated symbol \footnote{For a general compact group this would be (\ref{eq: conjugated clebsch-gordan}).}

    \begin{equation}
        \epsilon\left (\hat{\alpha}, \hat{\beta}, \hat{\gamma}\right )_{i j k} = \overline{\epsilon} \left (\hat{\alpha}, \hat{\beta}, \hat{\gamma}\right )_{i j k} \, .
    \end{equation}

    \item We can define a 1j symbol $\phi$ that transforms a representation into its complex conjugate as

    \begin{equation}
        R_{\hat{\alpha}} \left (g\right )_{i}^{\quad j} = \phi\left (\hat{\alpha}\right )^{\ast}_{i l} \, R^{\ast}_{\hat{\alpha}} \left (g\right )^{l}_{\quad m} \, \phi\left (\hat{\alpha}\right )_{jm} \, .
    \end{equation}

    \item The 1j symbol $\phi^{\ast}$ can then be used to lower indices and $\phi$ to raise them. 1j symbols obey unitarity conditions \footnote{For a general compact group the unitarity conditions hold up to a phase factor that is fixed to be $\pm 1$ (chapter 4 in \cite{Butler1975}).}

    \begin{equation}
    \phi^{\ast}\left (\hat{\alpha}\right )_{ij} \, \phi\left (\hat{\alpha}\right )_{i j'} = \delta_{j, j'} \quad ; \quad \phi^{\ast}\left (\hat{\alpha}\right )_{i' j} \phi\left (\hat{\alpha}\right )_{ij} = \delta_{i',i} \, .
    \label{eq: unitarity 1j}
    \end{equation}
    
\end{enumerate}

In principle raising or lowering indices could introduce phase factors in the form of 1j symbols. However, as long as we lower and raise at the same time two indices which are contracted, we do not get additional factors due to the unitarity conditions (\ref{eq: unitarity 1j}). \par
6j symbols depend on six representations, they are defined as

\begin{equation}
    \begin{pmatrix}
        \hat{\rho}_{1} & \hat{\rho}_{2} & \hat{\rho}_{3} \\
        \hat{\rho}_{4} & \hat{\rho}_{5} & \hat{\rho}_{6}
    \end{pmatrix} := \overline{\epsilon}\left (\hat{\rho}_{1}, \hat{\rho}_{2}, \hat{\rho}_{3}\right )_{k_{1} k_{2} k_{3}} \, \epsilon\left (\hat{\rho}_{1}, \hat{\rho}_{5}, \hat{\rho}_{6}\right )^{k_{1} \quad k_{6}}_{\quad k_{5}} \, \epsilon\left (\hat{\rho}_{4}, \hat{\rho}_{2}, \hat{\rho}_{6}\right )^{k_{4} k_{2}}_{\quad \quad k_{6}} \, \epsilon\left (\hat{\rho}_{4}, \hat{\rho}_{5}, \hat{\rho}_{3}\right )^{\quad k_{5} k_{3}}_{k_{4}} \, ,
\end{equation}
and they obey the following quadratic equation: the Racah back-coupling rule
\begin{equation}
    \sum_{\hat{\beta}_{3}} \dim\left (\hat{\beta}_{3}\right ) \left (-1\right )^{\hat{\beta}_{3} + \hat{\rho}_{4} + \hat{\rho}_{2}} \begin{pmatrix}
        \hat{\rho}_{3} & \hat{\alpha}_{2} & \hat{\rho}_{4} \\
        \hat{\alpha}_{1} & \hat{\rho}_{1} & \hat{\beta}_{3}
    \end{pmatrix}
    \begin{pmatrix}
        \hat{\alpha}_{2} & \hat{\alpha}_{1} & \hat{\beta}_{3} \\
        \hat{\rho}_{3} & \hat{\rho}_{1} & \hat{\rho}_{2}
    \end{pmatrix} =
    \begin{pmatrix}
        \hat{\alpha}_{2} & \hat{\rho}_{3} & \hat{\rho}_{4} \\
        \hat{\alpha}_{1} & \hat{\rho}_{1} & \hat{\rho}_{2} 
    \end{pmatrix} \, .
    \label{eq: 6j symbol identity}
\end{equation}

In section \ref{section: IRF crossing}, we perform computations for the IRF model only for simply reducible groups. However, we believe such computations still hold for general compact Lie groups. 6j symbols can be defined for any compact group (see eq. (9.6) in \cite{Butler1975}, eq. (5.1) in \cite{Derome} or the definition in \cite{Iraso:2018mwa}). A generalization of Racah back coupling rule also exists (see eq. (21) in \cite{Butler1975} or Theorem 5 in \cite{Derome}). In the general case there are more phase factors to keep track of, as we have highlighted in the footnotes. We also have to reintroduce the $\mu$ index and keep track of the conjugated representations.

\subsection{A sanity check}

For simply reducible groups like $SU\left (2\right )$ all fusion numbers are either $0$ or $1$, but in general fusion numbers can be greater than $1$ \cite{Alex:2010wi}. This apparently gives rise to some concern when performing the computations. Consider the expectation value of a Wilson loop $\hat{R}$ on a 2-sphere built by gluing two plaquettes with only one corner instead of three and with $B=1$. We get

\begin{equation}
\begin{split}
    Z_{\mathbb{S}^{2}} \left (\epsilon_{1}, \epsilon_{2}; \hat{R}\right ) =& \int dUdV \sum_{\hat{\sigma}, \hat{\rho}} \dim\left (\hat{\sigma}\right ) \dim\left (\hat{\rho}\right ) \chi_{\hat{\sigma}} \left (UV\right ) \chi_{\hat{\rho}}\left (U^{-1}V^{-1}\right ) \chi_{\hat{R}}\left (UV\right ) \times \\
    & \times e^{-\frac{\epsilon_{1}}{2} c_{2}\left (\hat{\rho}\right )} e^{-\frac{\epsilon_{2}}{2} c_{2}\left (\hat{\sigma}\right )} \\
    =& \sum_{\hat{\sigma}, \hat{\rho}, \mu, \nu} \dim\left (\hat{\sigma}\right ) \dim\left (\hat{\rho}\right ) \epsilon_{\mu}\left (\hat{\sigma}, \hat{R}, \overline{\hat{\rho}}\right )^{ae}_{d} \overline{\epsilon}_{\mu}\left (\hat{\sigma}, \hat{R}, \overline{\hat{\rho}}\right )^{c}_{bf} \epsilon_{\nu}\left (\hat{\sigma}, \hat{R}, \overline{\hat{\rho}}\right )^{bf}_{c} \times \\
    & \times \overline{\epsilon}_{\nu}\left (\hat{\sigma}, \hat{R}, \overline{\hat{\rho}}\right )^{d}_{ae} e^{-\frac{\epsilon_{1}}{2} c_{2}\left (\hat{\rho}\right )} e^{-\frac{\epsilon_{2}}{2} c_{2} \left(\hat{\sigma} \right )}\, .
\end{split} 
\label{eq: sum of CG coeff}
\end{equation}

\noindent We obtain two copies of 3j symbols instead of one. We might be concerned about getting $\left (N^{\hat{\rho}}_{\hat{\sigma} \hat{R}}\right )^{2}$, which would give a different result for the new plaquette, hence spoiling subdivision invariance. However, this is not the case. Using the properties of Haar measure (\ref{eq: Haar property 1} ) and (\ref{eq: normalisation haar measure}) we have

\begin{equation}
    \int dU dV \chi_{\hat{\sigma}}\left (UV \right) \chi_{\hat{\rho}} \left (U^{-1}V^{-1} \right) \chi_{\hat{R}} \left (UV \right) =
    \int dK dU \chi_{\hat{\sigma}}\left (K \right) \chi_{\hat{\rho}} \left (K^{-1} \right) \chi_{\hat{R}} \left (K \right) = N^{\hat{\rho}}_{\hat{\sigma} \hat{R}} \, , 
\end{equation}
where we defined $K=UV$. From this result we can deduce how to resolve the sums of coefficients appearing in (\ref{eq: sum of CG coeff})

\begin{equation}
\begin{split}      
    N^{\hat{\rho}}_{\hat{\sigma} \hat{R}}  =& \sum_{\mu, \nu} \delta_{\mu, \nu} \epsilon_{\mu} \left( \hat{\sigma}, \hat{R}, \overline{\hat{\rho}} \right
    )^{ae}_{d} \overline{\epsilon}_{\nu} \left( \hat{\sigma}, \hat{R}, \overline{\hat{\rho}} \right) ^{d}_{ae} \\
    =& \sum_{\mu, \nu} \epsilon_{\mu} \left( \hat{\sigma}, \hat{R}, \overline{\hat{\rho}} \right)^{ae}_{d} \overline{\epsilon}_{\mu}\left(\hat{\sigma}, \hat{R}, \overline{\hat{\rho}} \right)^{c}_{bf} \epsilon_{\nu}\left (\hat{\sigma}, \hat{R}, \overline{\hat{\rho}} \right)^{bf}_{c} \overline{\epsilon}_{\nu}\left(\hat{\sigma}, \hat{R}, \overline{\hat{\rho}}\right )^{d}_{ae} \, .
    \label{eq: CG contraction}
\end{split}
\end{equation}
In particular, the relation
\[\epsilon_{\nu} \left(\hat{\sigma}, \hat{R}, \overline{\hat{\rho}} \right)^{bf}_{c} \, \overline{\epsilon}_{\mu}\left (\hat{\sigma}, \hat{R}, \overline{\hat{\rho}} \right)^{c}_{bf} = \delta_{\mu, \nu} \, ,\]
is a special case of eq. (5.5) in \cite{Butler1975} with the $i_{3}$ index summed over so that gives a $\dim \left(\lambda_{3} \right)$ canceling the additional $\dim \left(\lambda_{3} \right)$ factor on the right hand side.

\section{Theta Terms}
\label{appendix: theta terms}

Let us summarize here a few key properties of theta terms in 2d gauge theories \cite{Sharpe:2014tca}. 
Theta terms appear as phases, corresponding to different weightings for the bundles summed over in the path integral.
In a model with a trivial theta angle, all bundles are weighted the same. \par
Isomorphism classes of $G$-bundles over closed surfaces with $G$ a connected simple compact Lie group are in one to one correspondence with

\begin{equation}
    H^{2}(\Sigma; \pi_{1}(G)) \simeq \pi_{1}(G) \, .
\end{equation}

As a consequence, the corresponding discrete theta angles take values in the space of unitary representations $\hat\theta \in \widehat{\pi_{1}(G)}$. 

Equivalently, a discrete theta angle is a choice of element in \cite{Sharpe:2014tca, Santilli:2024dyz}

\begin{equation}
    \Hom \left(\pi_{1} \left(G \right), U \left(1 \right) \right)^{\pi_{0} \left(G \right)} \, ,
\end{equation}
where the superscript means that the $\Hom$ space is restricted to the $\pi_{0} \left(G \right)$-invariant part. In or notation, $G$ is the quotient of its universal cover  $\tilde{G}$ by the discrete group $H$, hence $\pi_1\left(G\right)\simeq H$.
\par

\bibliographystyle{JHEP}
\bibliography{biblio}
\addcontentsline{toc}{section}{References}

\end{document}